\documentclass[twocolumn,aps,showpacs,prb,tightenlines,superscriptaddress,10pt]{revtex4-1}
\usepackage{graphicx}
\usepackage{amssymb}
\usepackage{dcolumn}
\usepackage{amsmath}
\usepackage{bm}
\usepackage{float}
\usepackage{hyperref}
\usepackage[section]{placeins}
\usepackage{epstopdf}

\begin{document}

\title{Normal and abnormal electron-hole pairs in a voltage-pulse-driven quantum conductor}
\author{X. K. Yue}
\author{Y. Yin}
\thanks{Author to  whom correspondence should be addressed}
\email{yin80@scu.edu.cn.}
\affiliation{College of Physical Science and Technology,
  Sichuan University, Chengdu, Sichuan, 610065, China}
\date{\today}

\begin{abstract}
  Electron-hole pairs can be excited coherently in a quantum conductor by applying voltage pulses on its contact. We
  find that these electron-hole pairs can be classified into two kinds, whose excitation probabilities have different
  dependence on the Faraday flux of the pulse. Most of the pairs are of the first kind, which can be referred to as
  ``normal'' pairs. Their excitation probabilities increase nearly monotonically with the flux and saturate to the
  maximum value $1$ when the flux is large enough. In contrast, there exist ``abnormal'' pairs, whose excitation
  probabilities can exhibit oscillations with the flux. These pairs can only be excited by pulses with small width. Due
  to the oscillation of the probabilities, the abnormal pairs can lead to different features in the full counting
  statistics of the electron-hole pairs for pulses with integer and noninteger fluxes.
\end{abstract}


\maketitle

\section{Introduction}
\label{sec1}

\begin{figure}
  \centering
  \includegraphics[width=6.5cm]{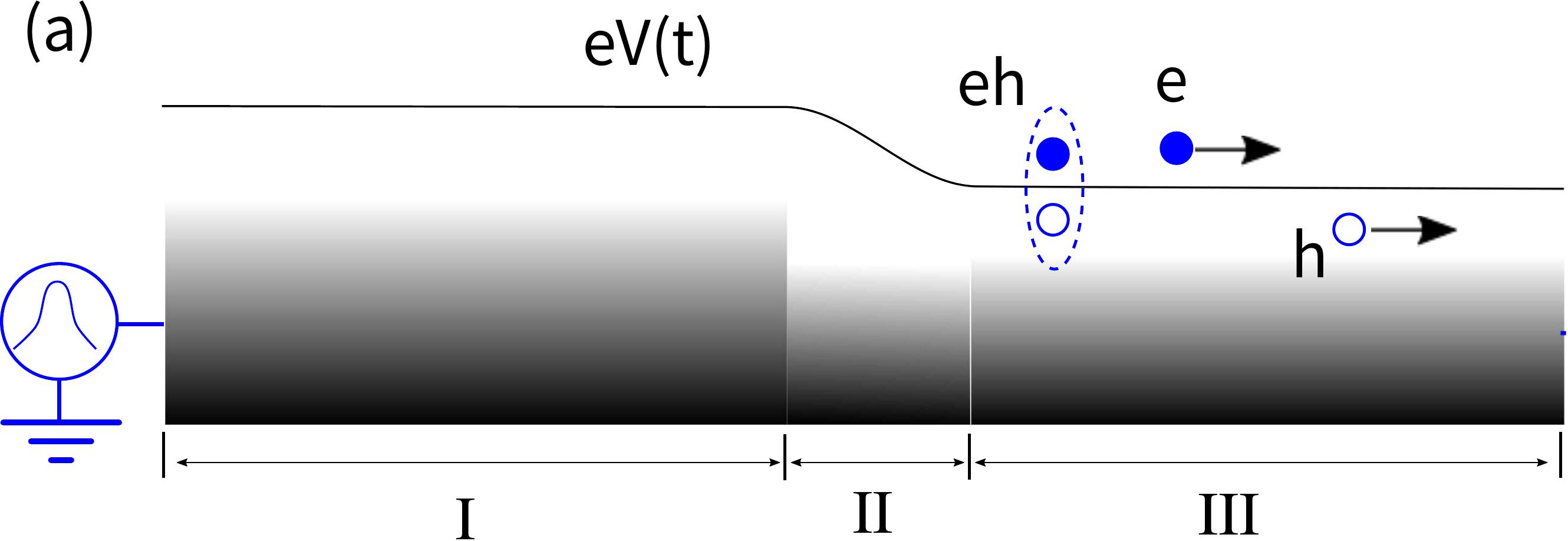}
  \includegraphics[width=6.5cm]{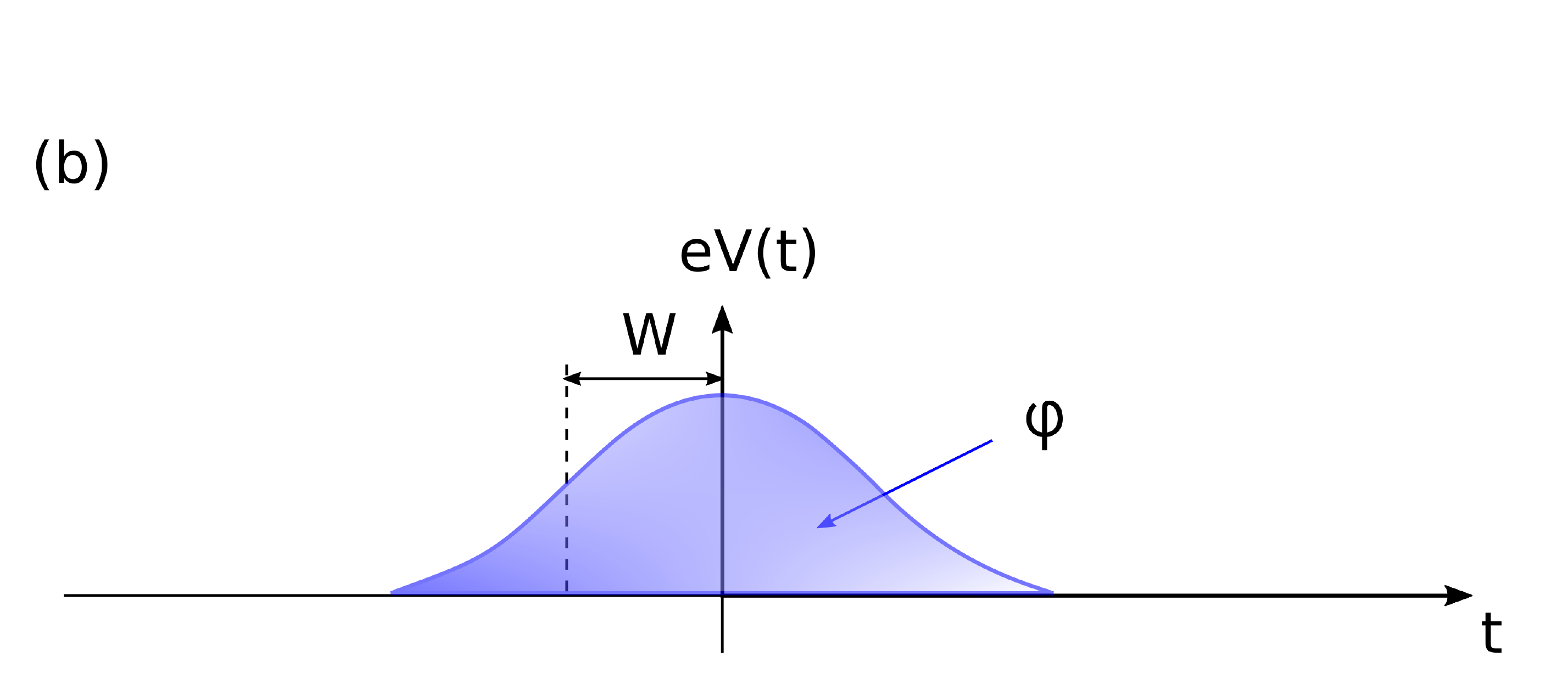}
  \includegraphics[width=6.5cm]{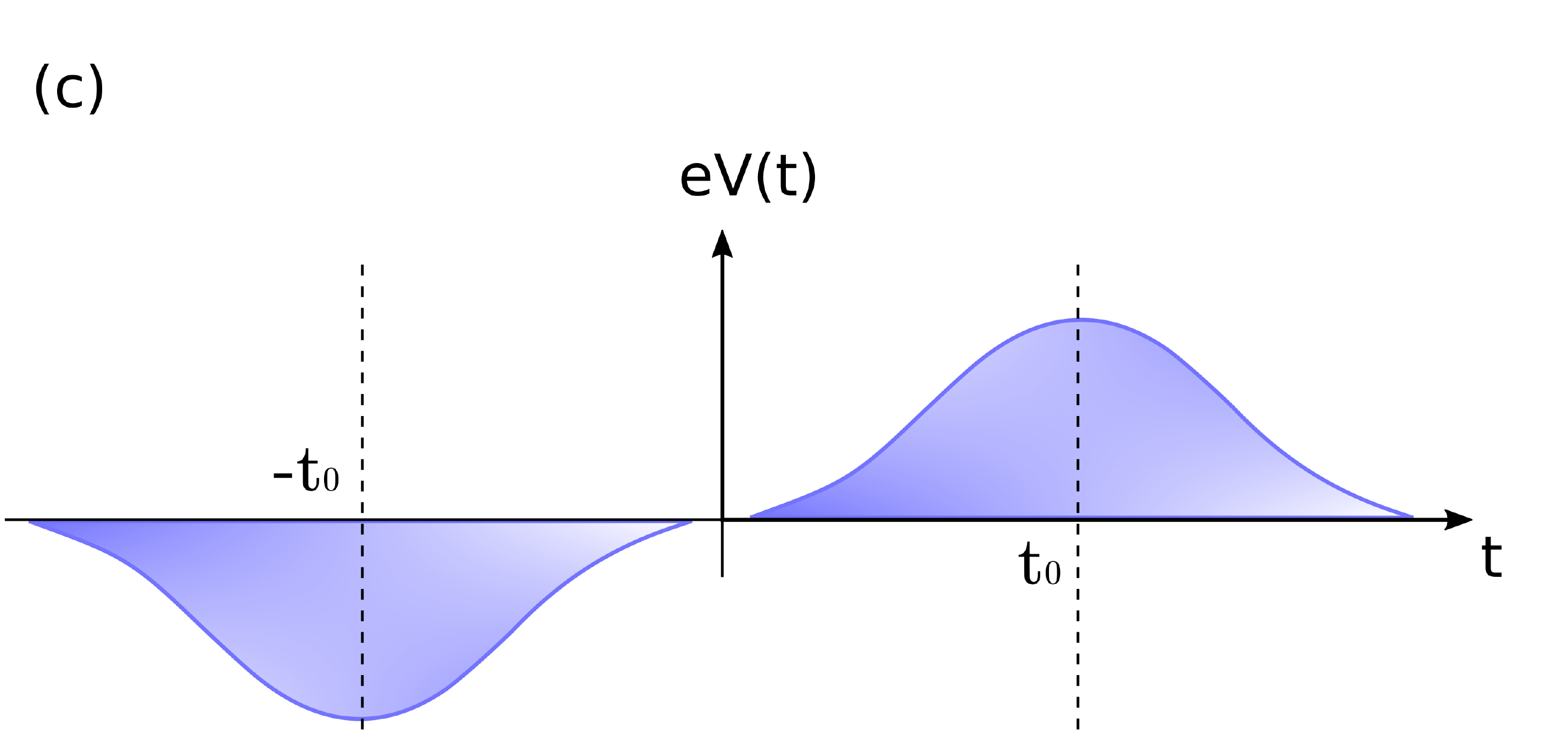}
  \caption{ (a) Schematic of the on-demand charge injection via a voltage pulse $V(t)$. By applying $V(t)$ on the
    contact of the quantum conductor, electrons (e) and/or holes (h) from the reservoir(region I) can be injected into
    the quantum conductor(region III), accompanied with additional electron-hole pairs (eh). The voltage drop is assumed
    to occur across a short interval at the interface(region II). (b) Schematic of the applied voltage pulse. The
    overall profile of the voltage pulse can be characterized by its half width at half maximum (HWHM) $W$ and Faraday
    flux $\varphi$. (c) Schematic of the voltage pulse for electron-hole pair excitation. The two pulses has the same
    shape but opposite signs.}
  \label{fig1}
\end{figure}

The on-demand coherent injection of single or few charges in solid state devices has attracted much attention in the
recent decade.\cite{keeling-2006-minim-excit, feve-2007-deman-coher, keeling-2008-coher-partic, mahe-2010-curren-correl,
  fletcher-2013-clock-contr, baeuerle-2018-coher-contr} In a simple way, such injection can be realized by applying a
nanosecond voltage pulse on the Ohmic contact of a quantum conductor at sub-kelvin temperatures, as illustrated in
Fig.~\ref{fig1}(a). The injected charges are carried by electrons and/or holes in the Fermi sea of the quantum
conductor,\cite{landau-1957-theor-fermi-liquid, pines-2018-theor-quant-liquid} whose quantum states are well-defined and
can be manipulated via the voltage pulse. This setup has been referred to as the voltage pulse electron source, which
offers a simple and feasible way to achieve the time-triggered coherent injection.\cite{glattli-2016-levit-elect} While
a negative pulse tends to inject electrons, a positive pulse tends to inject holes. However, additional
electron-hole(eh) pairs can also be excited during the injection, manifesting themselves in the noise of the injected
charges.\cite{levitov-1996-elect-count, vanevic-2007-elemen-event}

The statistics of the eh pairs show different features for pulses with integer and noninteger Faraday fluxes. To
minimized the noise, pulses with integer fluxes are favorable, since the excitation of the eh pairs are suppressed in
this case.\cite{levitov-1996-elect-count} Remarkably, the eh pairs can be totally eliminated when the pulse is further
tuned to be the form of the Lorentzian.\cite{ivanov-1997-coher-states} In doing so, one obtains a noiseless current
carried by only electrons or holes, whose wave function has a semi-exponential profile in the energy
domain.\cite{keeling-2006-minim-excit} They are now referred to as levitons, which play a central role on the on-demand
charge injection.\cite{dubois-2013-minim-excit, bocquillon-2014-elect-quant, glattli-2016-levit-elect}

In contrast, a large amount of eh pairs can be excited via pulses with noninteger fluxes. In fact, the number of eh
pairs detected over a large time interval $t$ diverges as $t$ increasing, which is closely related to the dynamical
orthogonality catastrophe.\cite{levitov-1996-elect-count, dubois-2013-integ-fract, glattli-2018-pseud-binar} In this
case, the quantum states of the eh pairs can show unique features, which can be seen from the interference pattern in
the Mach-Zehnder interferometers.\cite{hofer-2014-mach-zehnd} Moreover, new types of excitations can be constructed from
these states. For example, it has been proposed that, by applying a Lorentzian pulse with one-half flux, a zero-energy
quasiparticle with fractional charge $e/2$ can be created, which is described by a mixed state and cannot exist in the
absence of the eh pairs.\cite{moskalets-2016-fract-charg}

These studies suggest that the different statistics of the eh pairs can be attributed to their different quantum
states. To gain a more comprehensive understanding of such difference, explicit expressions of these quantum states are
favorable. However, they are only known for certain specific pulses.\cite{keeling-2006-minim-excit,
  vanevic-2016-elect-elect, yin-2019-quasip-states} A general description of these states for arbitrary pulses are still
missing, which hinder further developments along this direction.

As a first step toward solving this problem, in this paper we consider the case when two successive pulses with the same
shape but opposite signs are applied on the voltage pulse electron source, as illustrated in Fig.~\ref{fig1}(c). In this
case, the net injected charges are zero and only eh pairs are excited, making it easier to extract their
information. For each eh pair, we show that both the excitation probability and the one-body wave function can be
obtained from the corresponding scattering matrix of the electron source, which offer a comprehensive description of the
quantum states of the eh pairs.

By using such description, we show that all the eh pairs excited by the voltage pulse can be classified into two kinds,
despite the detailed shape of the pulse. The excitation probabilities of the two kinds of eh pairs exhibit quite
different dependence on the flux of the pulse: For the first kind of eh pairs, their probabilities increase nearly
monotonically with the flux. They can saturate to the maximum value $1$ when the flux is large enough. Most of the eh
pairs belong to this kind, which we refer to as ``normal'' eh pairs. In contrast, the probabilities for the second kind
of eh pairs undergo oscillations with the flux. These pairs can only be excited for pulses with small width, which we
here refer to as ``abnormal'' eh pairs.

We find that the abnormal pairs can play an important role on the full counting statistics (FCS) of the eh pairs. For
the voltage pulse electron source, we find that the corresponding FCS can be characterized by an effective binomial
distribution, whose cumulant generating function $S(\chi)$ has the form:
$S(\chi) = \alpha \ln( 1 - \bar{p}_k + e^{i \chi} \bar{p}_k)$, indicating that the electron source can excite
effectively $\alpha$ eh pairs with an effective probability $\bar{p}$. The parameter $\alpha$ {\em without} and {\em
  with} the contribution of the abnormal eh pairs can show qualitatively different behaviors as a function of the flux
$\varphi$: {\em Without} the contribution of the abnormal eh pairs, the parameter $\alpha$ exhibits a sequencing of
plateaus. The abnormal eh pairs can lead to a derivation from these plateaus, demonstrating the impact of the abnormal
eh pairs clearly.

The paper is organized as follows: In Sec.~\ref{sec2}, we present the model for the voltage pulse electron source and
show how to extract the quantum states of the eh pairs from the scattering matrix. In Sec.~\ref{sec3}, by using a
Gaussian-shaped pulse as an example, we show how to classify the two kinds of eh pairs from their excitation
probabilities. Their impact on the full counting statistics is also discussed in this section. In Sec.~\ref{sec4}, we
show that the two kinds of eh pairs can also be found for voltage pulses with different profiles. In particular, we
demonstrate how does the abnormal eh pairs evolve when the profile of the pulse approaches the Lorentzian. We summarized
in Sec.~\ref{sec5}.

\section{Model and formalism}
\label{sec2}

The voltage pulse electron source can be modeled as a single-mode quantum conductor, where a time-dependent voltage
$V(t)$ is applied on the Ohmic contact of the conductor, as illustrated in Fig.~\ref{fig1}(a). We assume that $V(t)$ has
the form in the time domain
\begin{eqnarray}
  V(t) & = & V_p(t_0+t) - V_p(t_0-t),
  \label{s2:eq1}
\end{eqnarray}
indicating that it is composed of two successive pulses[$\pm V_p(t)$] with the same shape but opposite signs, which are
separated by a time interval $t_0$. The width of each pulse can be characterized by the half width at half maximum
(HWHM) $W$, while the strength can be described by the Faraday flux
$\varphi = (e/h) \int^{+\infty}_{-\infty} V_p(t) dt$, as illustrated in Fig.~\ref{fig1}(b). The time interval $t_0$ is
usually chosen to be larger than the width $W$ of each pulse ($W < t_0$), so that the two pulses are well-separated in
time domain, as illustrated in Fig.~\ref{fig1}(c).

In the spatial domain, the voltage drop between the contact and the conductor is assumed to occur across a short
interval so that the corresponding dwell time $\tau_D$ satisfies: $ k_B T_e \ll \hbar/W \ll \hbar/\tau_D \ll E_F$, with
$E_F$ representing the Fermi energy and $T_e$ representing the electron temperature. In this case, the corresponding
scattering matrix in the energy domain can be written as\cite{beenakker-2005-optim-spin}
\begin{eqnarray}
  b(E) = \int^{+\infty}_{-\infty} \frac{d{E'}}{2\pi\hbar} S(E - E') a(E'),
  \label{s2:eq1-1}
\end{eqnarray}
where $a(E)$ and $b(E)$ represent the electron annihilation operators in the Ohmic contact and the quantum conductor,
respectively. The matrix element $S(E - E')$ is only the function of the energy difference $E-E'$, which can be related
to the voltage pulse $V(t)$ as
\begin{eqnarray}
  S(E - E') = \int^{+\infty}_{-\infty} dt e^{ \frac{i}{\hbar} [(E-E') t - e \int^t d\tau V(\tau)] }.
  \label{s2:eq1-2}
\end{eqnarray}
Note that for the voltage pulse we considered here, the scattering matrix is symmetric, {\em i.e.},
$S(E, E') = S(E', E)$, since we have $V(t) = -V(-t)$ from Eq.~\eqref{s2:eq1}.

It is convenient to write the scattering matrix into the form of the polar decomposition in the energy
domain.\cite{beenakker-1997-random-matrix, jalabert-2000-new-direc, mello-2004-quant-trans} For the symmetric scattering
matrix, the decomposition can be written as\cite{yin-2019-quasip-states}
\begin{eqnarray}
  S(E, E') & = & \sum_k \left[ \begin{tabular}{cc}
                                 $\psi^e_k(E)$, & $\psi^h_k(E)$\\
                               \end{tabular}\right]^{\ast} \nonumber\\
           &&\hspace{-1cm} \times \left[\begin{tabular}{cc}
                                          $\sqrt{ 1 - p_k }$ & $i \sqrt{p_k}$\\
                                          $i \sqrt{p_k}$ & $\sqrt{ 1 - p_k}$\\
                                        \end{tabular}\right] \left[\begin{tabular}{c}
                                                                     $\psi^e_k(E')$\\
                                                                     $\psi^h_k(E')$\\
                                                                   \end{tabular}\right],
  \label{s2:eq2}
\end{eqnarray}
with $k$ being a positive integer and the notation ${...}^{\ast}$ denoting the complex conjunction. The quantity $p_k$
is real and lies in the region $[0, 1]$. The two functions $\psi^e_k(E)$ and $\psi^h_k(E)$ are both complex and satisfy:
$\psi^e_k(E) = 0$ for $E \le 0$ and $\psi^h_k(E) = 0$ for $E > 0$.

Given the scattering matrix, one can construct the many-body state of the quantum conductor from the first-order
electronic correlation function via the Bloch-Messiah reduction.\cite{yin-2019-quasip-states} In doing so, one obtains
the many-body state in the zero-temperature limit($T_e \to 0$) as
\begin{eqnarray}
  \hspace{-0.5cm} | \Psi_b \rangle & = & \sum_k \Big[ \sqrt{1 - p_k} + i \sqrt{p_k} B^{\dagger}_e(k) B^{\dagger}_h(k) \Big] | F \rangle, \label{s2:eq3}
\end{eqnarray}
where $|F\rangle$ represents the Fermi sea and $B^{\dagger}_e(k)$[$B^{\dagger}_h(k)$] represents the creation operator
for the electron[hole] component. They can be related to the polar decomposition Eq.~\eqref{s2:eq2} as
\begin{eqnarray}
  B^{\dagger}_e(k) & = & \int^{+\infty}_0 \frac{dE}{2\pi\hbar} \psi^e_k(E) a^{\dagger}(E), \nonumber\\
  B^{\dagger}_h(k) & = & \int^0_{-\infty} \frac{dE}{2\pi\hbar} \psi^h_k(E) a(E),
                         \label{s2:eq3-1}
\end{eqnarray}
This form suggests that only eh pairs are excited in the Fermi sea by the voltage pulses $V(t)$. The quantum state of
each pair are characterized by the excitation probability $p_k$ and the one-body wave function of the electron[hole]
component $\psi^{e[h]}_k(E)$, which can be solely decided from the polar decomposition of the scattering matrix shown in
Eq.~\eqref{s2:eq2}.

The above equations establish a general relation between the voltage pulse $V(t)$ and the quantum states of the eh
pairs. By choosing $t_0$ as the time unit, the overall shape of the pulse can be characterized by the (dimensionless)
width $W/t_0$ and the flux $\varphi$, while the fine structure of the shape is decided by the detailed profile of
$V_p(t)$. All these three ingredients can affect the quantum state of the eh pairs.

\section{Classification of the electron-hole pairs}
\label{sec3}

Despite the different profile of $V_p(t)$, we find that the eh pairs can always be classified into two kinds, which can
be seen from their excitation probabilities $p_k$. To demonstrate this, let us consider the case of the Gaussian
profile, when $V_p(t)$ has the form
\begin{eqnarray}
  V_p(t) & = &  \frac{2 \hbar \sqrt{\pi \ln{2}}}{e} \frac{\varphi}{W} \exp[-(\frac{t \sqrt{\ln{2}} }{W})^2].
               \label{s3:eq1}
\end{eqnarray}

\subsection{Excitation probability}

\begin{figure}
  \centering
  \includegraphics[width=6.5cm]{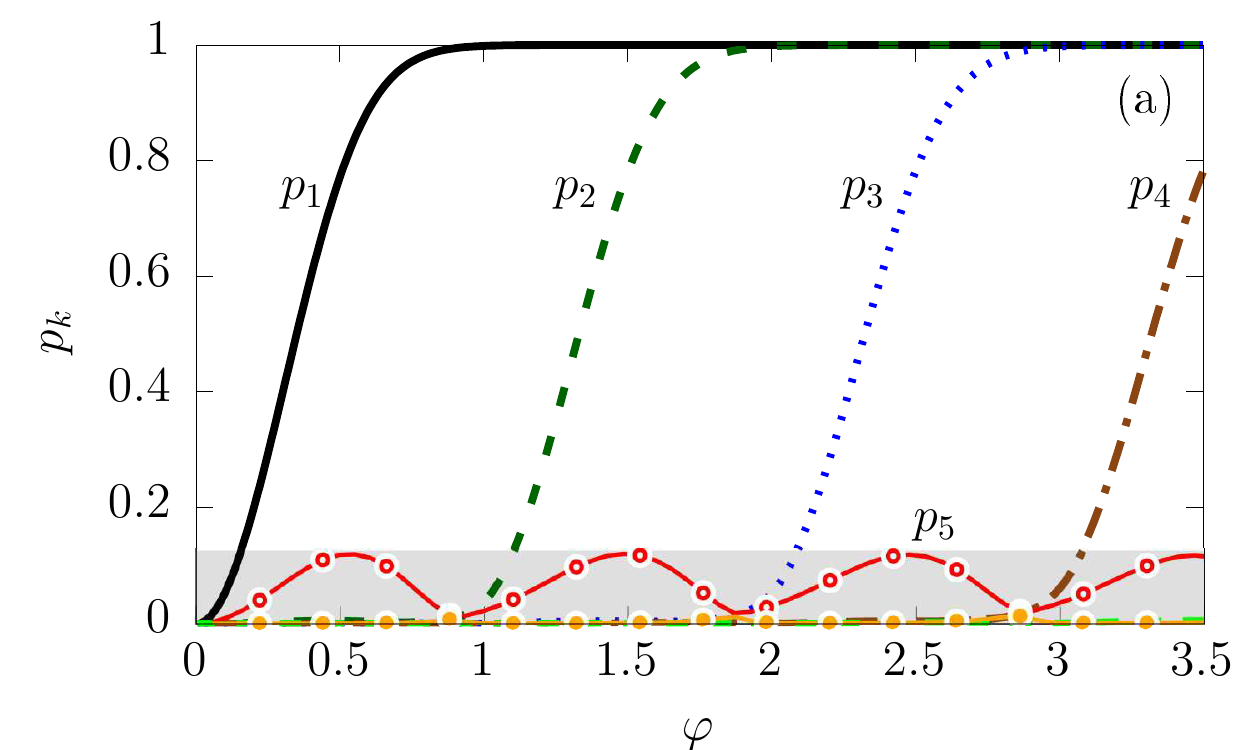}
  \includegraphics[width=6.5cm]{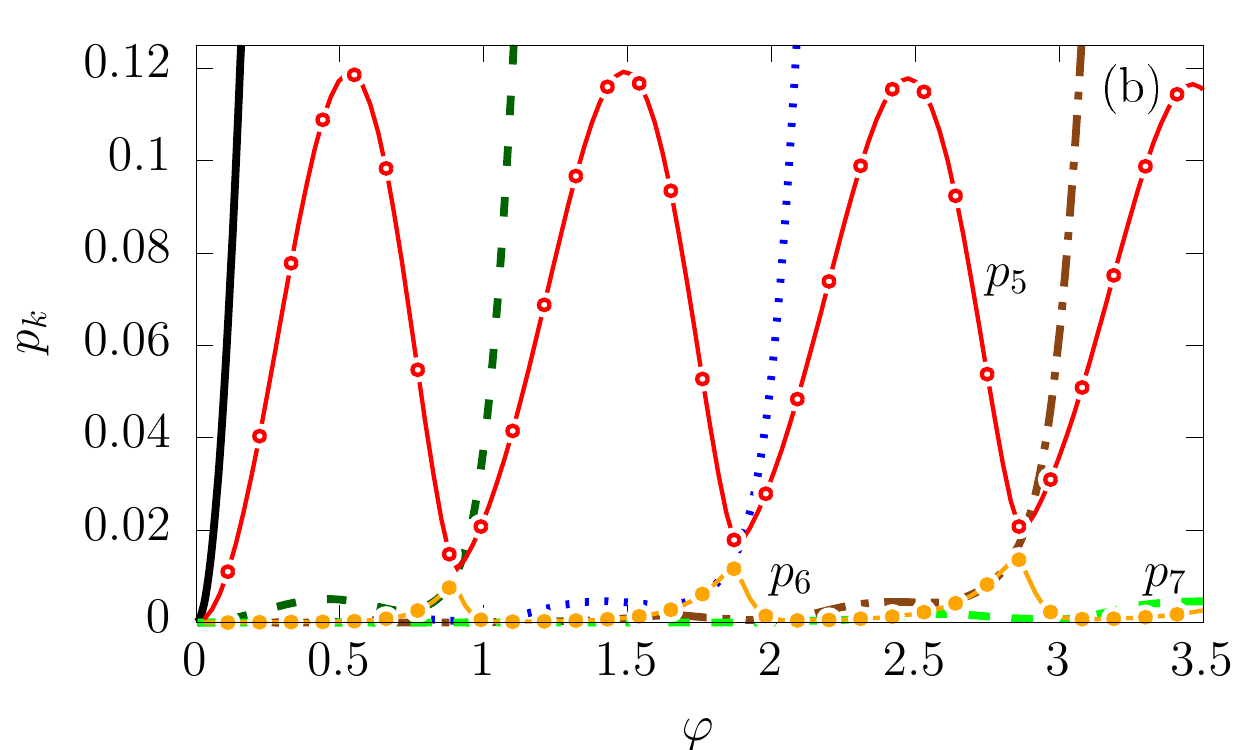}
  \caption{(Color online) (a) Excitation probabilities as functions of the flux $\varphi$ in the case of the Gaussian
    profile, corresponding to the width $W/t_0=1/8$. The black solid, green dashed, blue dotted and brown dash-dotted
    curves correspond to the first four normal eh pairs ($k=1$-$4$). The red solid curve with circles corresponds to the
    first abnormal eh pair ($k=5$). (b) Zoom-in plot of the grey regime. The orange solid curve with dots corresponds
    the second abnormal eh pair ($k=6$), while the bright-green dashed curve corresponds to the fifth normal eh pair
    ($k=7$). The minimums of the probability $p_5$ occur at $\varphi= 0.90$, $1.88$ and $2.87$; while the maximums of
    the probability $p_6$ occur at $\varphi= 0.89$, $1.87$ and $2.85$.}
  \label{fig2}
\end{figure}

We show the typical behavior of the probabilities $p_k$ as functions of the flux $\varphi$ in Fig.~\ref{fig2}(a),
corresponding to $W/t_0 = 1/8$.\footnote{The eh pairs can be distinguished via their one-body wave function. In this
  paper, we assume that the wave function of each eh pair evolves continuously with the flux $\varphi$, {\em i.e.},
  $\lim_{\delta \varphi \to 0} \int dE [\psi^{e(h)}_k(E; \varphi)]^{\ast} \psi^{e(h)}_{k'}(E; \varphi+\delta \varphi) =
  \delta_{k-k'}$. This allows us to study the $\varphi$-dependence of the probability for each eh pair individually.}
One can see that the excitation is dominated by the first five eh pairs ($k=1$-$5$). There are two additional eh pairs
($k=6$ and $7$), whose excitation probabilities are much smaller and can only be distinguished from the zoom-in plot
shown in Fig.~\ref{fig2}(b). The probabilities of the other eh pairs are negligible due to their small probabilities:
They are all smaller than $0.002$, which cannot be seen even in the zoom-in plot.

All these eh pairs can be classified into two kinds, whose probabilities exhibit quite different dependence on the flux
$\varphi$. There are five eh pairs ($k=1$-$4$ and $7$), which belong to the first kind. The corresponding probability
$p_k$ remains quite small when the flux $\varphi$ is below a certain threshold value. Above the threshold, $p_k$
increases monotonically and saturates to the maximum value $1$ when the flux $\varphi$ is large enough. For example, the
probability $p_2$ (green dashed curve) is kept below $0.04$ for $\varphi$ below the threshold $1.0$, which can only be
seen clearly from the zoom-in plot Fig.~\ref{fig2}(b). Note that in this regime, $p_2$ can change non-monotonically upon
the flux $\varphi$. For $\varphi$ above the threshold $1.0$, $p_2$ increases monotonically. It can reach above $0.99$
for $\varphi > 1.87$, as shown in Fig.~\ref{fig2}(a). It is worth noting that the similar behavior of the probabilities
has been reported in the case of the ac driving.\cite{vanevic-2008-elemen-charg, yin-2019-quasip-states} This is easy to
be understood, since the voltage pulse $V(t)$ we studied here [see, Fig~\ref{fig1}(c)] can be regarded as a single
period of ac driving voltage. Due to such similarity in the probabilities, we here refer to the eh pairs of the first
kind as {\em ``normal''} eh pairs.

In contrast, for eh pairs of the second kind ($k=5$ and $6$), the corresponding probabilities undergo oscillations with
the flux $\varphi$. The probability $p_5$ exhibits minimums around the point when $\varphi$ takes integer values, while
$p_6$ tends to exhibit maximums at almost the same positions [see the caption in Fig.~\ref{fig2}(b) for the detailed
positions]. This kind of eh pairs is absent in the case of the ac driving, hence we here refer to them as {\em
  ``abnormal''} eh pairs.

\begin{figure}
  \centering
  \includegraphics[width=6.5cm]{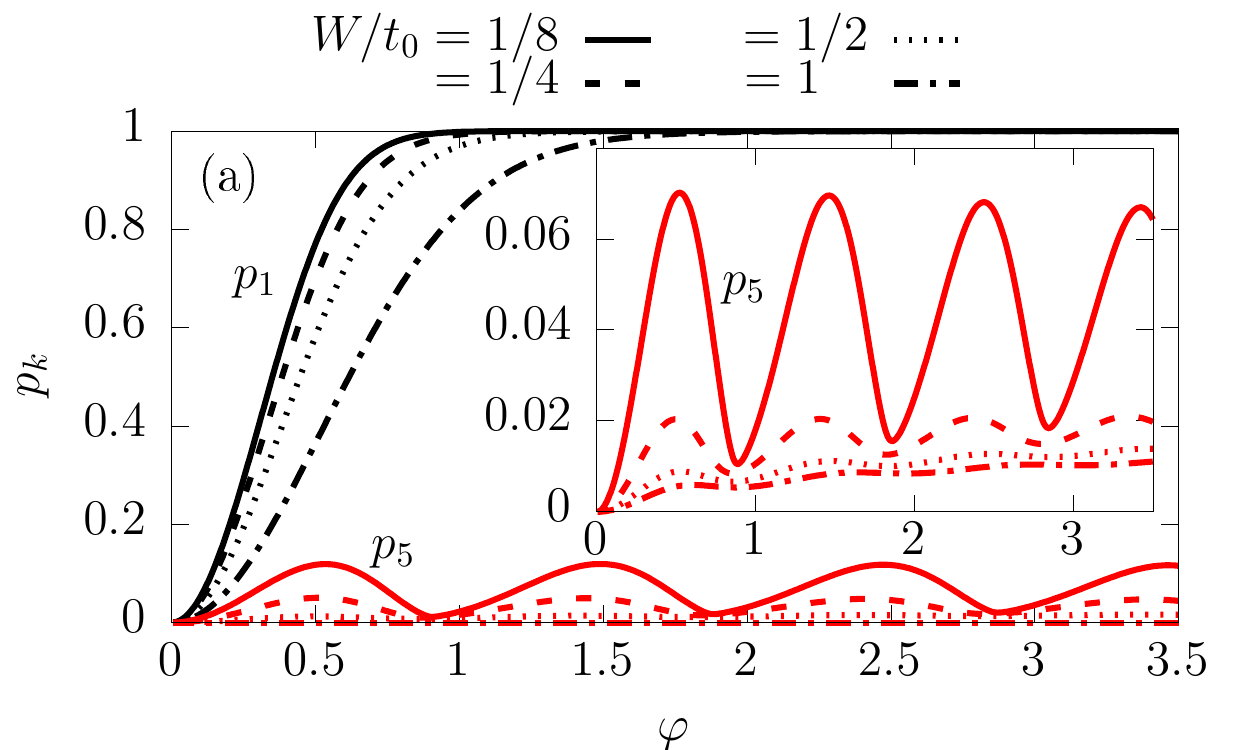}
  \includegraphics[width=6.5cm]{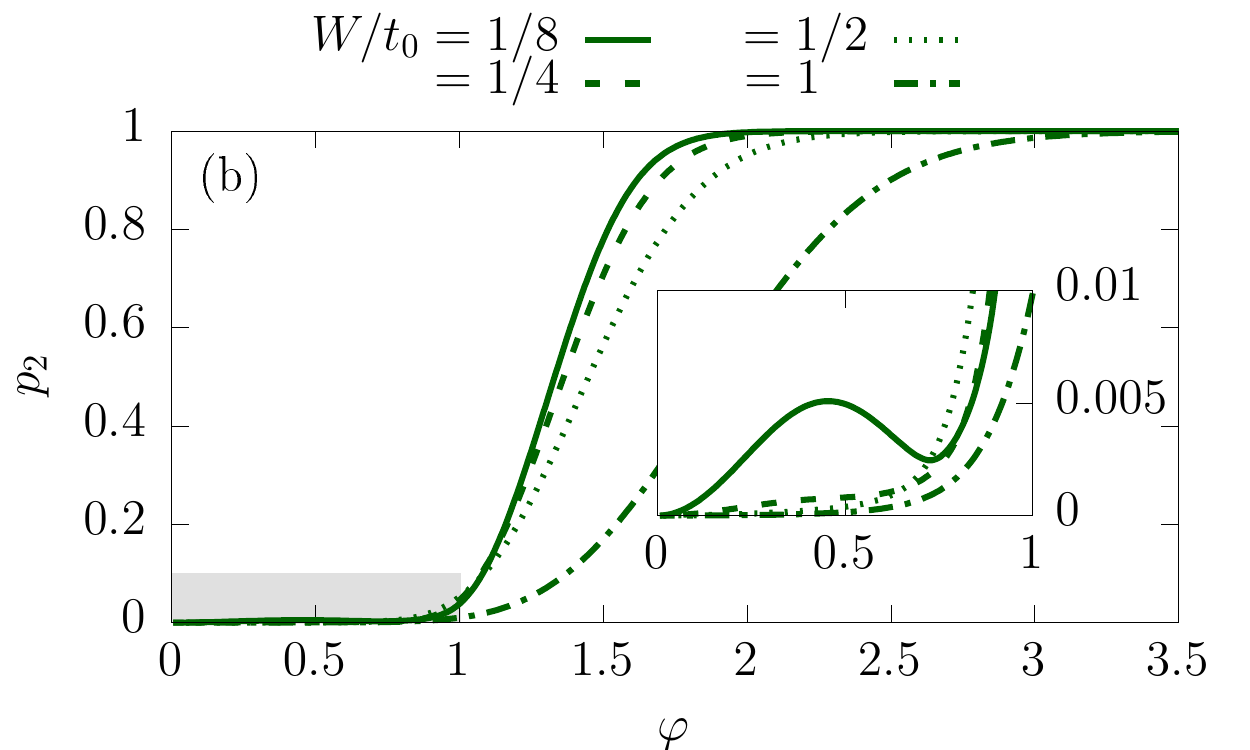}
  \caption{(Color online) (a) The probabilities $p_1$ (black curve) and $p_5$ (red curve) as functions of the flux
    $\varphi$ for different pulse width $W/t_0$. The detailed behavior of $p_5$ is shown in the inset. (b) The
    probability $p_2$ as functions of the flux $\varphi$ for different pulse width $W/t_0$. The inset shows the zoom-in
    plot of the grey regime.}
  \label{fig3}
\end{figure}

One may wonder why the abnormal eh pairs cannot be excited via the ac driving voltage? The main reason is that the width
$W/t_0$ corresponding to a single period of typical ac driving is rather large ($W/t_0 \sim 1$). In this case, the
excitation of the abnormal eh pairs is strongly suppressed. To illustrate this, we compare the probabilities $p_1$
(normal eh pair) and $p_5$ (abnormal eh pair) as functions of the flux $\varphi$ for different width $W/t_0$ in
Fig.~\ref{fig3}(a). The solid, dashed, dotted and dash-dotted curves correspond to the width $W/t_0=1/8$, $1/4$, $1/2$
and $1$, respectively. One can see that by increasing the width $W/t_0$, both the probabilities can be
suppressed. However, the suppression for $p_1$ is relatively weak and becomes marginal for $\varphi > 1.5$. So it can
still play an import role even for $W/t_0 = 1$. In contrast, the suppression for $p_5$ is much more pronounced. For
$W/t_0 \ge 1/2$, $p_5$ is too small and can only be seen clearly from the inset.

Note that the non-monotonically behavior of the normal eh pairs can also be suppressed as the pulse width $W/t_0$
increasing. This can be seen from Fig.~\ref{fig3}(b), where we plot the probability $p_2$ as functions of the flux
$\varphi$ with different width $W/t_0$. By comparing to $p_1$ shown in Fig.~\ref{fig3}(a), one can see that the
probability $p_2$ has a more sensitive dependence on $W/t_0$. Moreover, from the inset of Fig.~\ref{fig3}(b), one finds
that the non-monotonically behavior of $p_2$ can only be seen for $W/t_0=1/8$, as illustrated by the green solid
curve. For $W/t_0 \ge 1/4$, $p_2$ always increases monotonically as $\varphi$ increasing.

In theory, as the pulse width $W/t_0$ further decreasing, the abnormal eh pairs can play a more and more important
role. In the meantime, the non-monotonically behavior of the normal eh pairs can also be more pronounced. However, it is
difficult to realize a well-behaved nanosecond voltage pulse for too small width $W/t_0$ in practical. Experimentally,
one usually stays in the region for $W/t_0 > 0.1$.\cite{dubois-2013-integ-fract} In this region, the excitation
probabilities of the abnormal eh pairs are typically much smaller than the ones of the normal eh pairs. In fact, due to
the small probability $p_6$ [see Fig.~\ref{fig2}], the impact of the second abnormal eh pair ($k=6$) is negligible in
most cases and only the first one ($k=5$) is relevant. In the meantime, the non-monotonically behavior of the normal eh
pairs can usually play negligible roles, as the probability of the normal eh pairs are rather small below the
threshold.

\subsection{Full counting statistics}

Due to the oscillation of the probabilities, it is expected that the abnormal eh pairs can lead to different statistics
for pulses with integer and noninteger fluxes. To study this, we calculate the full counting statistics (FCS) $P(N)$ of
the eh pairs, corresponding to the probability of exciting $N$ eh pairs over the time interval $[-t_f/2, t_f/2]$. In the
limit $t_f \to +\infty$, the cumulant generating function (CGF) [$e^{S(\chi)} = \sum_N P(N) e^{i N \chi}$] can be
related to the excitation probability $p_k$ as\cite{yin-2019-quasip-states}
\begin{eqnarray}
  S(\chi) & = & \sum_k \ln( 1 - p_k + e^{i \chi} p_k).
                \label{s3-2:eq1-1}
\end{eqnarray}
This corresponds to the Poisson binomial distribution, which is typical for non-interacting
particles.\cite{levitov-1996-elect-count} It allows us to separate the contribution of the abnormal eh pairs from the
normal ones, making it easier to study their impacts.

\begin{figure}
  \centering
  \includegraphics[width=7.5cm]{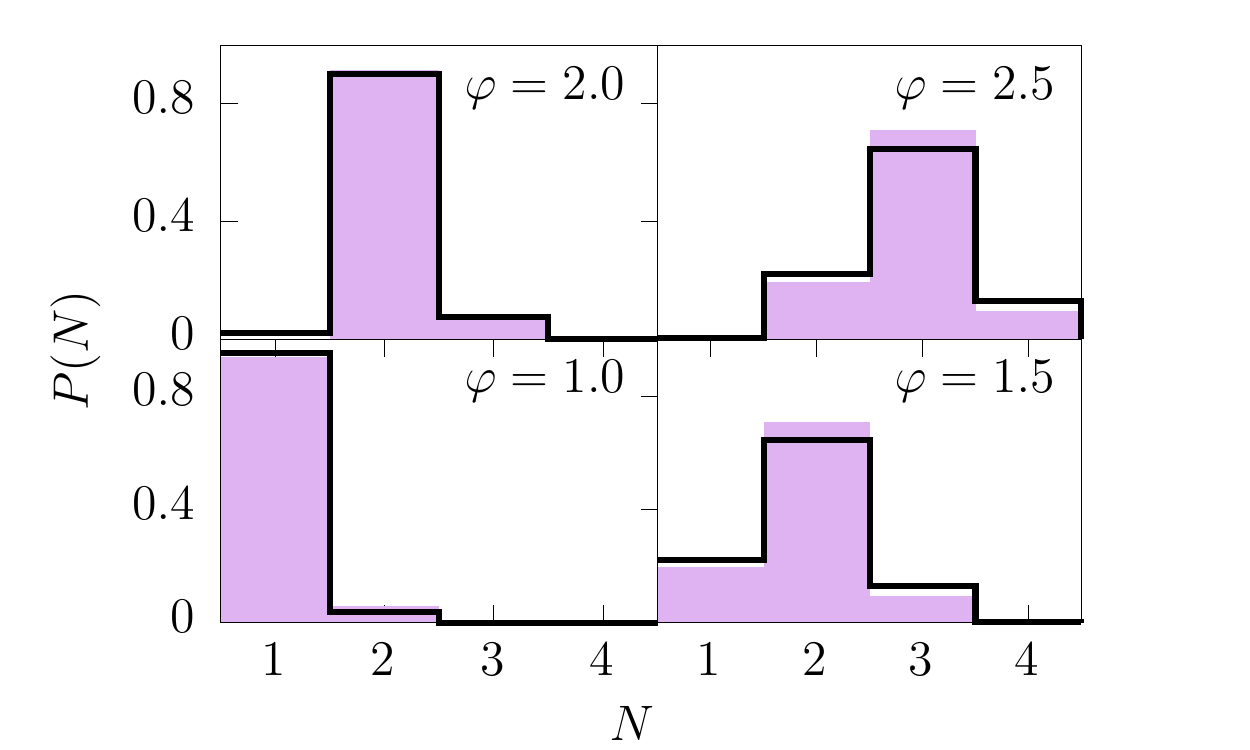}
  \caption{(Color online) The exact FCS P(N) from Eq.~\eqref{s3-2:eq1-1} (pink bars) and the FCS from the effective
    binomial distribution (black curves) with different pulse fluxes $\varphi=1.0$, $1.5$, $2.0$ and $2.5$. All the FCS
    are obtained by using the Gaussian profile with width $W/t_0=1/8$.}
  \label{fig4}
\end{figure}

Usually, the FCS are characterized by the mean $N_{ph} = \sum_N P(N) N$, variance
$\sigma^2_{ph} = \sum_N P(N) N^2 - N^2_{ph}$ and high-order cumulants, which can be obtained from derivatives of
$S(\chi)$. For the voltage pulse electron source we considered here, there exists an alternative way to characterize the
FCS. This is because the CGF in this case can be approximate as
\begin{eqnarray}
  S(\chi) & \approx & \alpha \ln( 1 - \bar{p} + e^{i \chi} \bar{p} ),
                      \label{s3-2:eq1-2}
\end{eqnarray}
corresponding to an effective binomial distribution. The two parameters $\alpha$ and $\bar{p}$ can be determined by
requiring the effective binomial distribution has the same mean and variance as the Poisson binomial distribution, {\em
  i.e.},
\begin{eqnarray}
  \alpha \bar{p} & = N_{ph} = & \sum_k p_k, \nonumber\\
  \alpha \bar{p} ( 1 - \bar{p} ) & = \sigma^2_{ph} = & \sum_k p_k (1 - p_k).
                      \label{s3-2:eq1-3}
\end{eqnarray}
In Fig.~\ref{fig4}, we compare the $P(N)$ from Eq.~\eqref{s3-2:eq1-1} (pink bars) with the $P(N)$ from
Eq.~\eqref{s3-2:eq1-2} (black curves) in several typical cases, which shows that the effective binomial distribution can
offer a good estimation of the overall behavior of the FCS.\footnote{Strictly speaking, the effective binomial
  distribution Eq.~\eqref{s3-2:eq1-2} cannot give a positive-defined $P(N)$ via the usual formula
  $P(N) = \int^{\pi}_{-\pi} \frac{d \chi}{2\pi} e^{S(\chi)-i N \chi}$, when the parameter $\alpha$ is not an integer. Here
  the FCS $P(N)$ corresponding to the effective binomial distribution is obtained via the saddle point
  approximation.\cite{yin-2019-quasip-states}.}

Hence the two parameters $\alpha$ and $\bar{p}$ can be used to characterize the FCS of the eh pairs. They indicate that
the voltage pulse can excite effectively $\alpha$ eh pairs with an effective probability $\bar{p}$, offering an
alternative but more intuitive way to interpret the physical meaning of the FCS. In particular, $\alpha$ and $\bar{p}$
can show different behaviors {\em without} and {\em with} the contribution of the abnormal eh pairs, from which their
impact on the FCS can be seen more clearly.

\begin{figure}
  \centering
  \includegraphics[width=6.5cm]{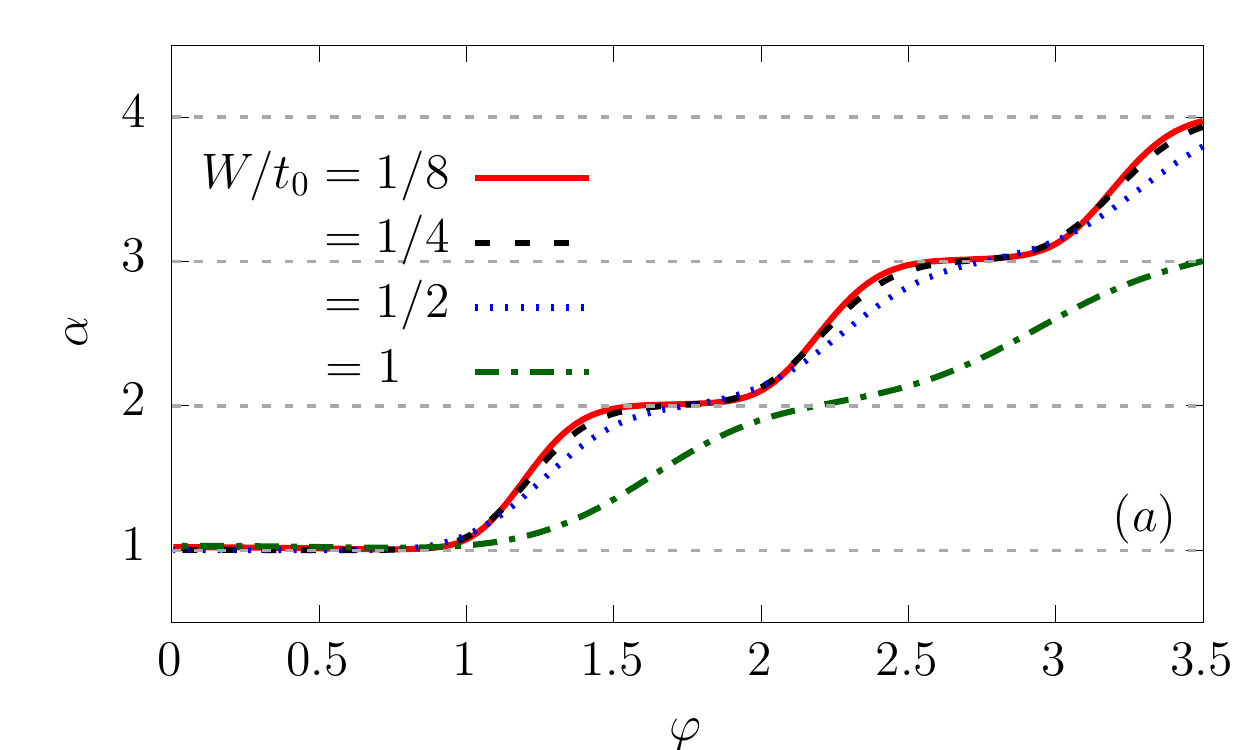}
  \includegraphics[width=6.5cm]{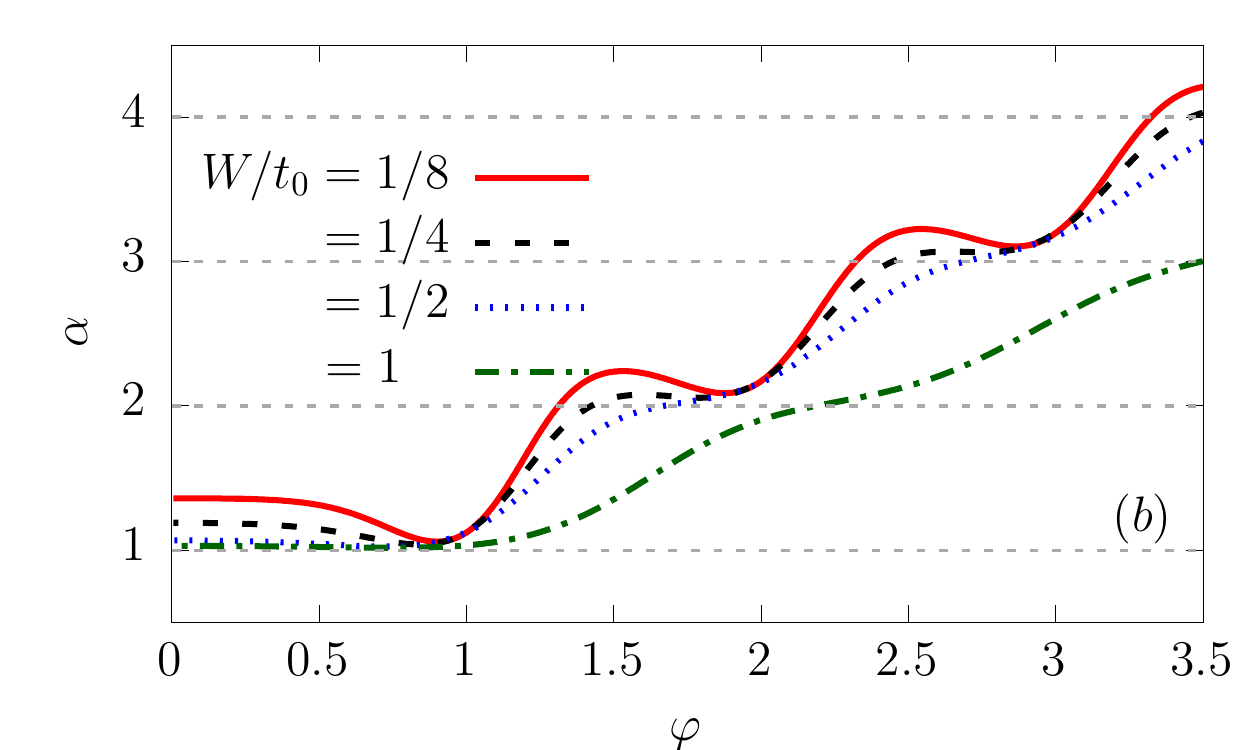}
  \caption{(Color online) The parameter $\alpha$ of the effective binomial distribution (a) {\em without} and (b) {\em
      with} the contribution of the abnormal eh pairs. The red solid, black dashed, blue dotted and green dash-dotted
    curves correspond to the width $W/t_0=1/8$, $1/4$, $1/2$ and $1$, respectively.}
  \label{fig5}
\end{figure}

To see this, let us first concentrate on the parameter $\alpha$ of the effective binomial distribution. We show $\alpha$
as a function of the flux $\varphi$ {\em without} and {\em with} the contribution of the abnormal eh pairs in
Fig.~\ref{fig5}(a) and (b), respectively. In the figure, different curves correspond to different pulse width $W/t_0$.
From Fig.~\ref{fig5}(a), one can see that, {\em without} the contribution of the abnormal eh pairs, $\alpha$ can exhibit
a sequencing of plateaus. The structure of the plateaus can be seen more clearly from the red solid curve, corresponding
to $W/t_0 = 1/8$. This curve shows that the plateaus are quantized at positive integer values $n=1, 2, 3, ...$, as
indicated by the grey dotted horizontal lines. As $\varphi$ increasing, $\alpha$ can change from the $n$th plateau to
the next one, whenever $\varphi$ is large than the corresponding integer $n$.

The plateaus are pronounced for small width $W/t_0$. As $W/t_0$ increasing, all the plateaus tend to diminish, except
for the lowest one. This can be seen by comparing the red solid curve to the black dashed ($W/t_0=1/4$), blue dotted
($W/t_0=1/2$) and green dash-dotted ($W/t_0=1$) ones in Fig.~\ref{fig5}(a). For $W/t_0 = 1$, only the lowest plateau
preserves, while the other plateau are merged into a smooth rise with small ripples.

The abnormal eh pairs can induce a derivation of $\alpha$ from the plateaus, as shown in Fig.~\ref{fig5}(b). Such
derivation can be seen more clearly by comparing the red solid curves in Fig.~\ref{fig5}(a) and (b), corresponding to
$W/t_0=1/8$. By comparing the two curves, one can see that the derivation is mainly due to the enhancement of $\alpha$
at noninteger fluxes. By further comparing them to the excitation probabilities in Fig.~\ref{fig2}, one can see that
such enhancement can be attributed to the contribution of the probability $p_5$: The enhancement is strong whenever
$p_5$ tends to exhibit a maximum. The only exception occurs around the point $\varphi=0$, where the enhancement of
$\alpha$ is rather strong, while the probability $p_5$ is dropping to zero. The main reason of such exception is that:
The excitation probabilities of the normal eh pairs also drop rapidly to zero as $\varphi \to 0$. This makes $p_5$ can
still lead to a significant contribution to the FCS at this point.

As the width $W/t_0$ increasing, $p_5$ decreases rapidly, as we have shown in Fig.~\ref{fig3}. Accordingly, the
enhancement of $\alpha$ becomes less and less pronounced, as can be seen in Fig.~\ref{fig5}(b). For $W/t_0 = 1/2$ (blue
dotted curve), the enhancement is too weak so that the derivation of $\alpha$ from the plateaus can no longer be
observed in the figure.

\begin{figure}
  \centering
  \includegraphics[width=6.5cm]{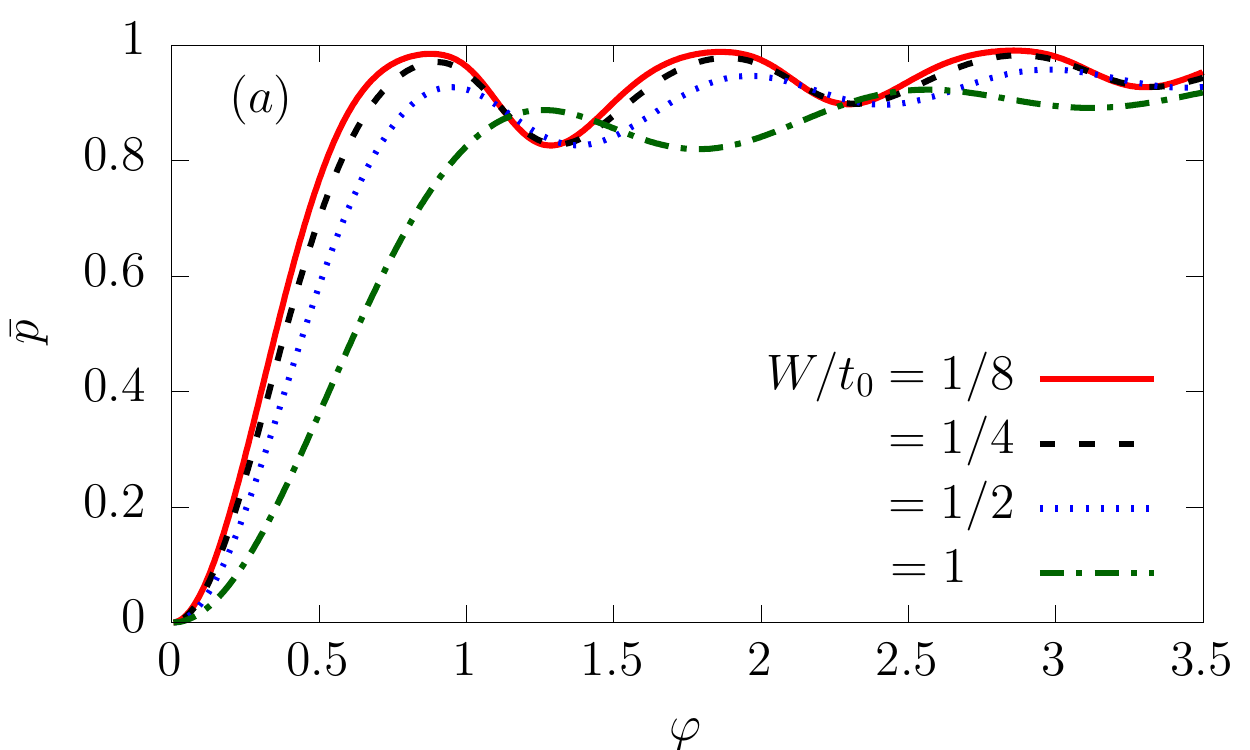}
  \includegraphics[width=6.5cm]{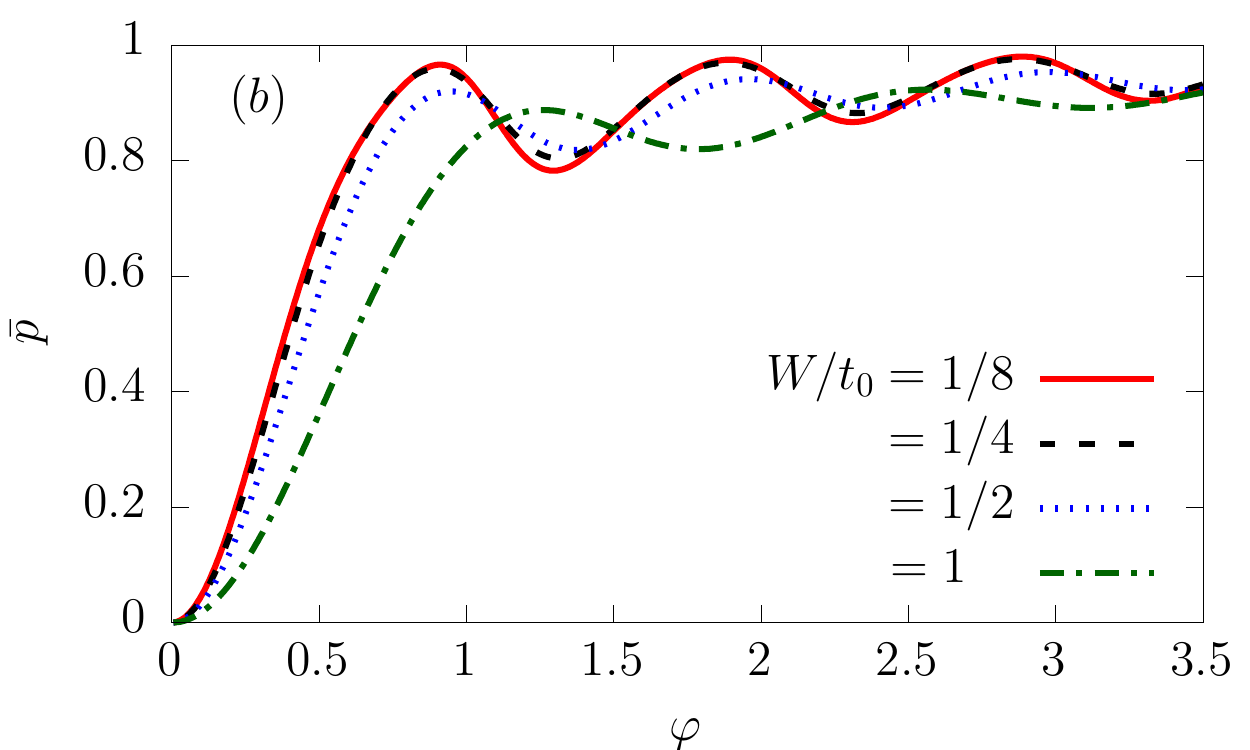}
  \caption{(Color online) The parameter $\bar{p}$ of the effective binomial distribution (a) {\em without} and (b) {\em
      with} the contribution of the abnormal eh pairs. The red solid, black dashed, blue dotted and green dash-dotted
    curves correspond to the width $W/t_0=1/8$, $1/4$, $1/2$ and $1$, respectively.}
  \label{fig6}
\end{figure}

While the abnormal eh pairs can affect the behavior of $\alpha$ distinctly, their impact on $\bar{p}$ is much less
pronounced. In fact, the probability $\bar{p}$ {\em without} and {\em with} the contribution of the abnormal eh pairs
show quite similar oscillations, which can be seen from Fig.~\ref{fig6}(a) and (b).

\section{Pulse shape dependence}
\label{sec4}

Although the above results are obtained by using the voltage pulse with the Gaussian profile, the classification of the
normal and abnormal eh pairs are quite general and can be seen for pulses with different profiles. We have checked the
excitation probabilities of the Square, Triangular and Parabolic profiles, which show quite similar behaviors as the
ones of the Gaussian profile [see Appendix~\ref{app1} for details]. Among various profiles, of particular interest is
the Lorentzian profile, which plays an important role in the study of levitons. One may wonder what happens to the
abnormal eh pairs when the profile of the pulse is tuned to be the Lorentzian. To demonstrate this, let us consider the
case of a mixed Lorentzian-Gaussian profile, when the corresponding $V_p(t)$ has the form
\begin{eqnarray}
  V_p(t) & = & r \frac{\hbar}{e} \frac{2 W \varphi}{ W^2 + t^2} \nonumber\\
         && + (1-r) \frac{\sqrt{2\pi}\hbar}{e} \frac{\varphi}{W} \exp[-(\frac{t \sqrt{\ln{2}}}{W})^2].
            \label{s4:eq1}
\end{eqnarray}
Here, the first term represents the Lorentzian profile, while the second term represents the Gaussian profile. The
parameter $r$ characterizes the degree of mixture between the two profiles. By increasing $r$ from $0$ to $1$, the
profile can evolve continuously from the Gaussian to the Lorentzian.

\begin{figure}
  \centering
  \includegraphics[width=6.5cm]{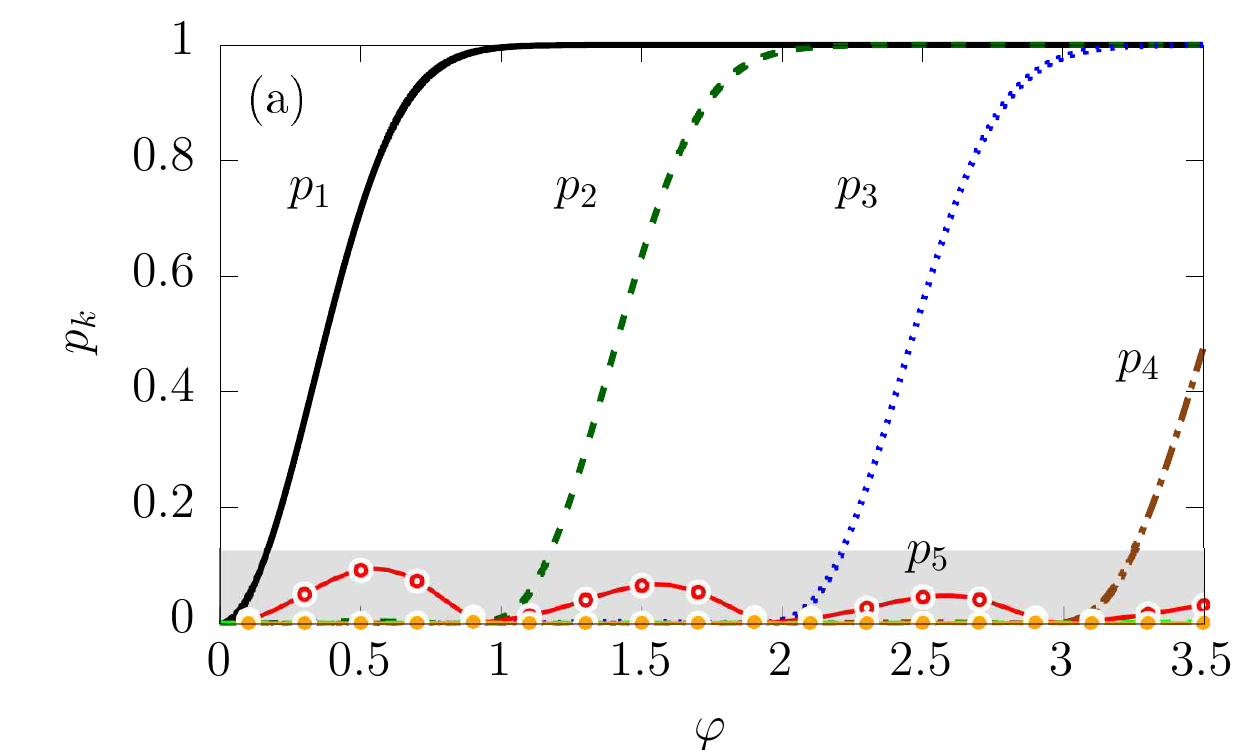}
  \includegraphics[width=6.5cm]{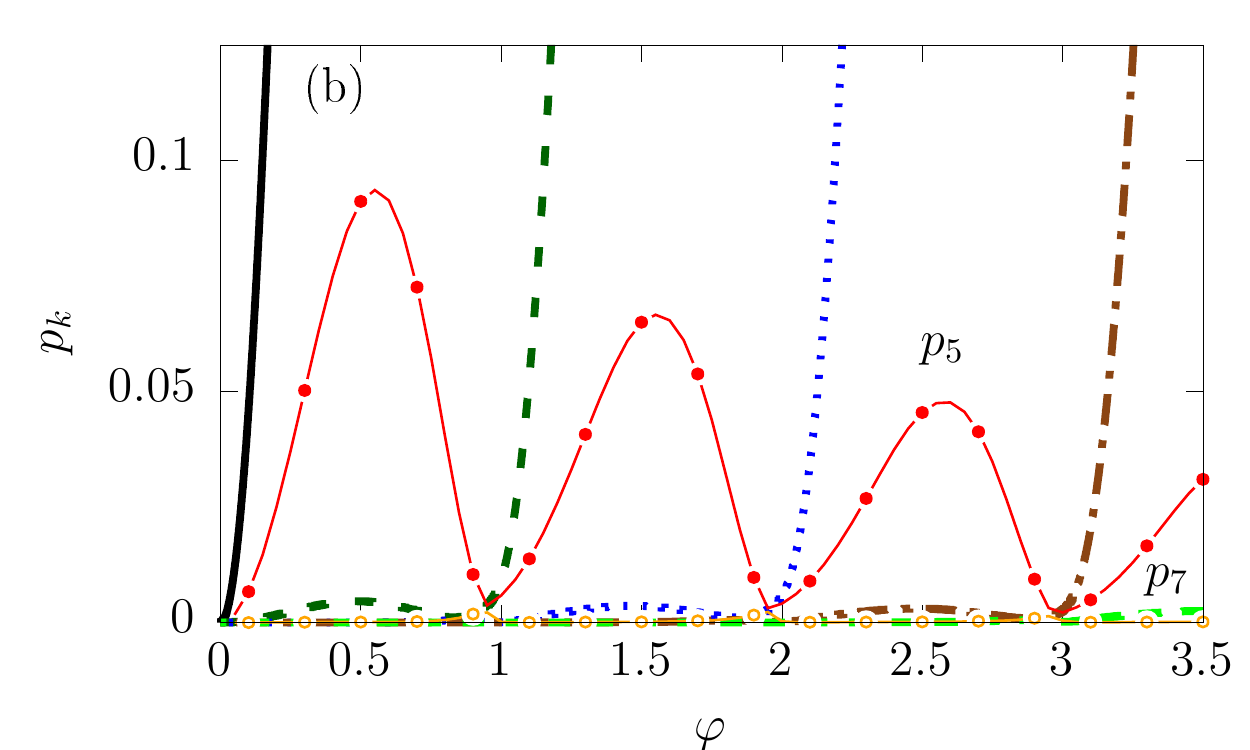}
  \caption{(Color online) The same as Fig.~\ref{fig2}, but with the mixed Lorentzian-Gaussian profile, where $r=0.4$ and
    $W/t_0=1/8$. Note that the probability $p_6$ of the second abnormal eh pair becomes too small to be seen, even in
    the zoom-in plot shown in (b).}
  \label{fig7}
\end{figure}

We show the typical behavior of the excitation probabilities as functions of the flux $\varphi$ in Fig.~\ref{fig7},
corresponding to $r=0.4$ and $W/t_0=1/8$. Comparing to Fig.~\ref{fig2} (corresponding to $r=0.0$), one can see that the
probabilities exhibit qualitatively similar behaviors, from which one can identify the normal and abnormal eh pairs
following the same procedure introduced in the previous section. Note that in this case, all the probabilities are
suppressed compared to the case of the Gaussian profile. For the normal eh pairs, the suppression is modest: One can
still find five normal ($k=1$-$4$ and $7$) eh pairs in this case, whose excitation probabilities exhibit quite similar
features as the ones of the Gaussian profile. In contrast, the suppression is more pronounced for the abnormal eh
pairs. Due to the suppression, only the excitation probability $p_5$ can be identified, corresponding to the first
abnormal eh pair. The probability $p_6$ of the second abnormal eh pair becomes too small to be observable, even in the
zoom-in plot shown in Fig.~\ref{fig7}(b).

\begin{figure}
  \centering
  \includegraphics[width=6.5cm]{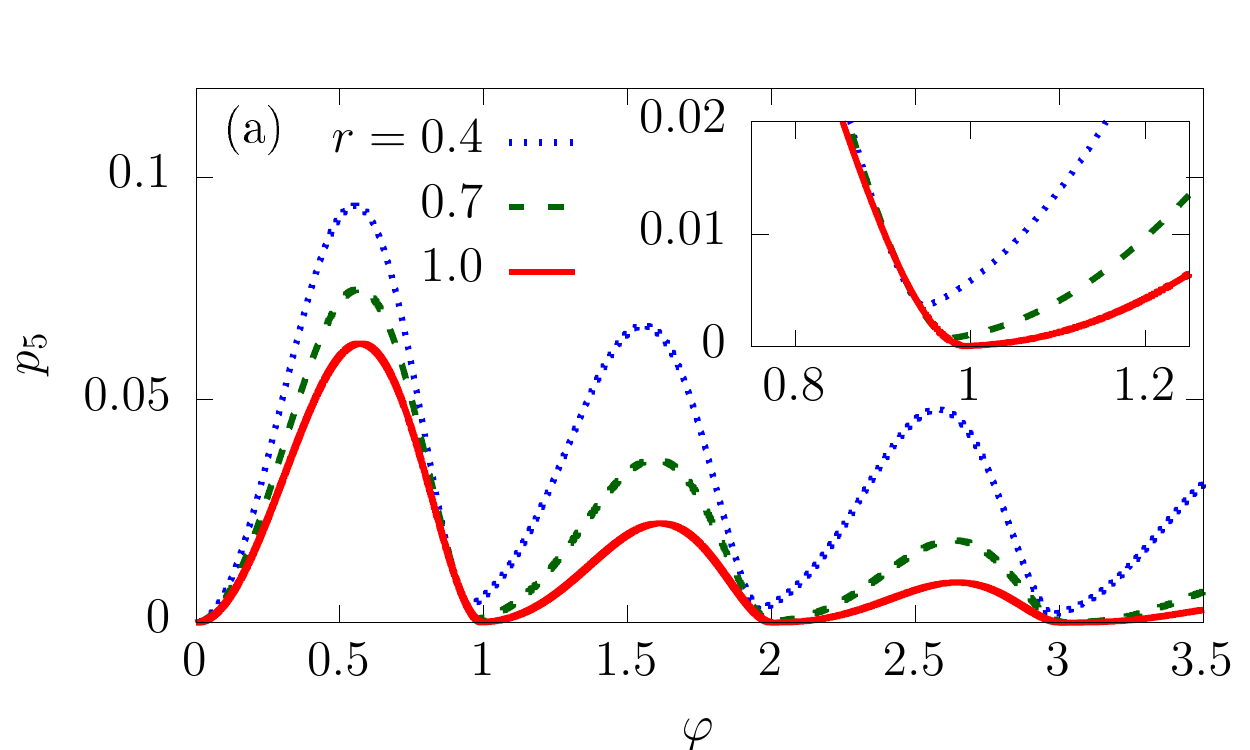}
  \includegraphics[width=6.5cm]{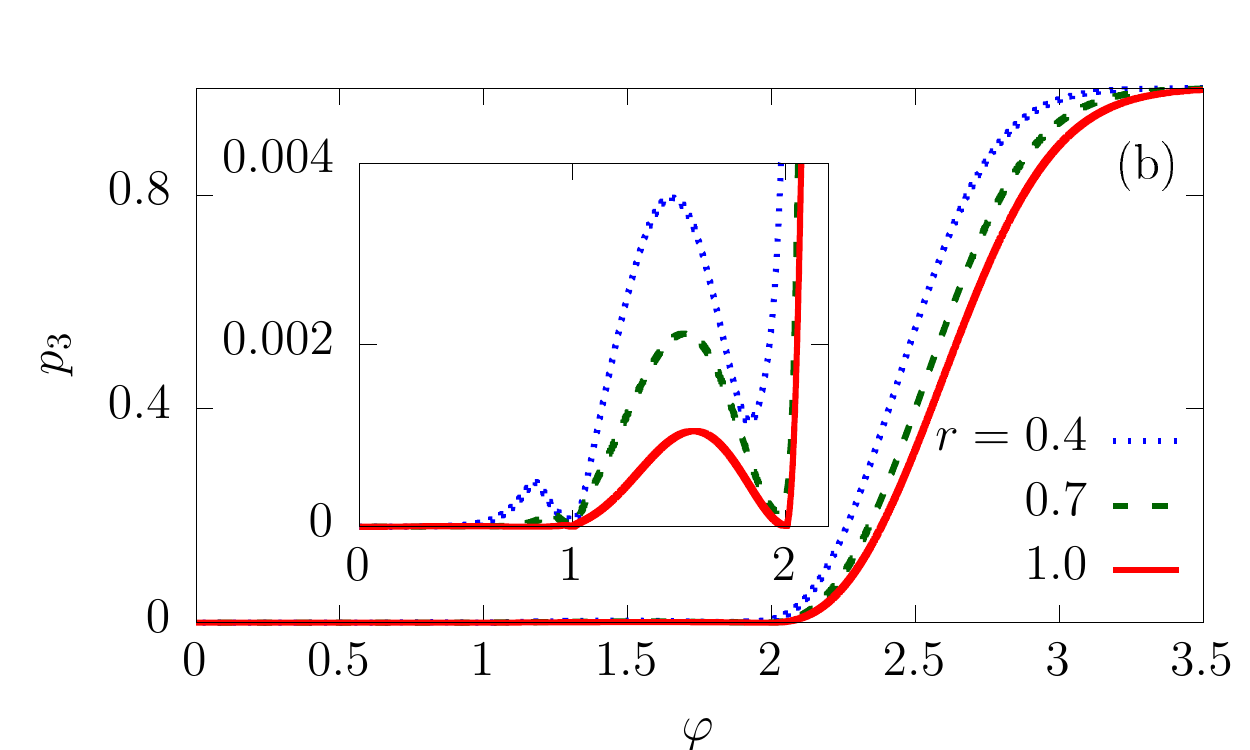}
  \caption{(Color online) (a) The probability $p_5$ as functions of the flux $\varphi$ with different mixture parameters
    $r$. The detailed behavior of $p_5$ around $\varphi=1$ is shown in the inset. (b) The probability $p_3$ as functions
    of the flux $\varphi$ with different mixture parameters $r$. The detailed behavior of $p_3$ below the threshold
    $2.0$ is shown in the inset.}
  \label{fig8}
\end{figure}

The suppression of the probability of the abnormal eh pairs is particular pronounced when the flux $\varphi$ takes
integer values. To clarify this, we show the probability $p_5$ with different mixture parameters $r$ in
Fig.~\ref{fig8}(a), corresponding to the first abnormal eh pair. One can see that the oscillation of $p_5$ undergoes a
damping oscillation with the flux $\varphi$. The damping becomes more and more strong as $r$ increasing. In the
meantime, the minimums of the oscillation move toward the points where $\varphi$ takes integer values. As $r$ reaching
$1.0$, all the minimums drop to zero at these points, as illustrated by the red solid curve in the figure. This feature
indicates that the excitation of the abnormal eh pairs is absent for the Lorentzian-shape pulse with integer fluxes.

The suppression of the probabilities at integer fluxes can also be seen from the normal eh pairs, but in a more subtle
way. To demonstrate this, we plot the probability $p_3$ with different mixture parameters $r$ in Fig.~\ref{fig8}(b),
corresponding to the third normal eh pair. For $r=0.4$ (blue dotted curve), one can see that $p_3$ is kept smaller than
$0.004$ below the threshold $\varphi = 2.0$, which can be seen more clearly from the inset. For $\varphi > 2.0$, $p_3$
increase monotonically and can reach almost the maximum value $1$ when $\varphi = 3.5$. By increasing the mixture
parameter $r$ from $0.4$ to $1.0$, $p_3$ can be suppressed to zero for $\varphi=1.0$ and $2.0$, which are below the
threshold [see the inset]. For $\varphi > 2.0$, the suppression is absent and the probabilities $p_3$ for different $r$
show quite similar behaviors, as shown in the main panel of the Fig.~\ref{fig8}(b).

\section{Summary}
\label{sec5}

In this paper, we study the quantum states of the eh pairs excited in a voltage-pulse-driven quantum conductor. By using
the Gaussian-shaped pulse as an example, we show that the eh pairs can always be classified into the normal and abnormal
eh pairs, whose excitation probabilities exhibit different dependence on the flux of the pulse $\varphi$. For the normal
eh pairs, the probabilities increase nearly monotonically with the flux. They can reach the maximum value $1$ when the
flux is strong enough. In contrast, the excitation probabilities of the abnormal eh pairs undergo oscillation with the
flux. These pairs can only be excited for pulses with small width. In practical cases, only the first abnormal eh pair
is relevant.

We find that the abnormal eh pairs can lead to different features in the FCS of the eh pairs for pulses with integer and
noninteger fluxes. This features can be better seen from the effective binomial distribution, whose CGF has the form
$S(\chi) = \alpha \ln(1-\bar{p}+e^{i\chi}\bar{p})$. Without the contribution of the abnormal eh pairs, the parameter
$\alpha$ exhibits a sequencing of plateaus. The abnormal eh pairs can lead to a derivation from these plateaus, which
can be treated as a signature of the abnormal eh pairs.

We also find that the classification of the normal and abnormal eh pairs is quite general and can be found for pulses
with different profiles. In particular, as the profile of the pulse evolves from the Gaussian to the Lorentzian, we show
that the excitation of the abnormal eh pairs can be totally suppressed when the pulse flux $\varphi$ takes integer
values.

\begin{acknowledgments}
  This work was supported by National Key Basic Research Program of China under Grant No. 2016YFF0200403.
\end{acknowledgments}

\appendix

\section{Excitation probabilities of other profiles}
\label{app1}
In this appendix, we present the excitation probabilities for the Square, Triangular and Parabolic profiles:

Square profile:
\begin{eqnarray}
 V_{p}(t) =
  \begin{cases}
    \frac{\pi\hbar}{e} \frac{\varphi}{W}       &, \quad |t|<W \\
    0  &,  \quad \text{otherwise.}
  \end{cases}
  \label{A1:eq1}
\end{eqnarray}

\begin{figure}[H]
  \centering
  \includegraphics[width=6.5cm]{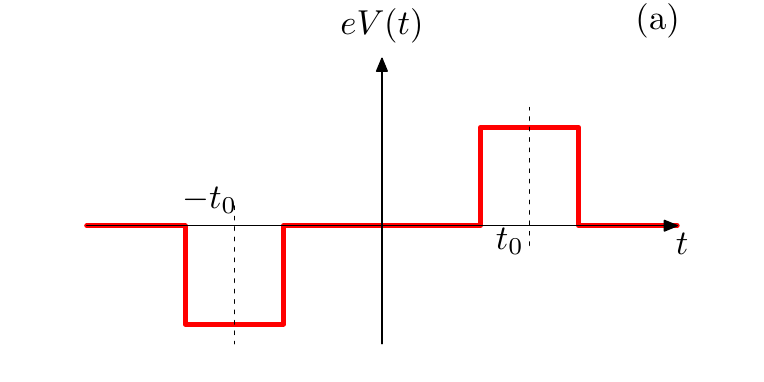}
  \includegraphics[width=6.5cm]{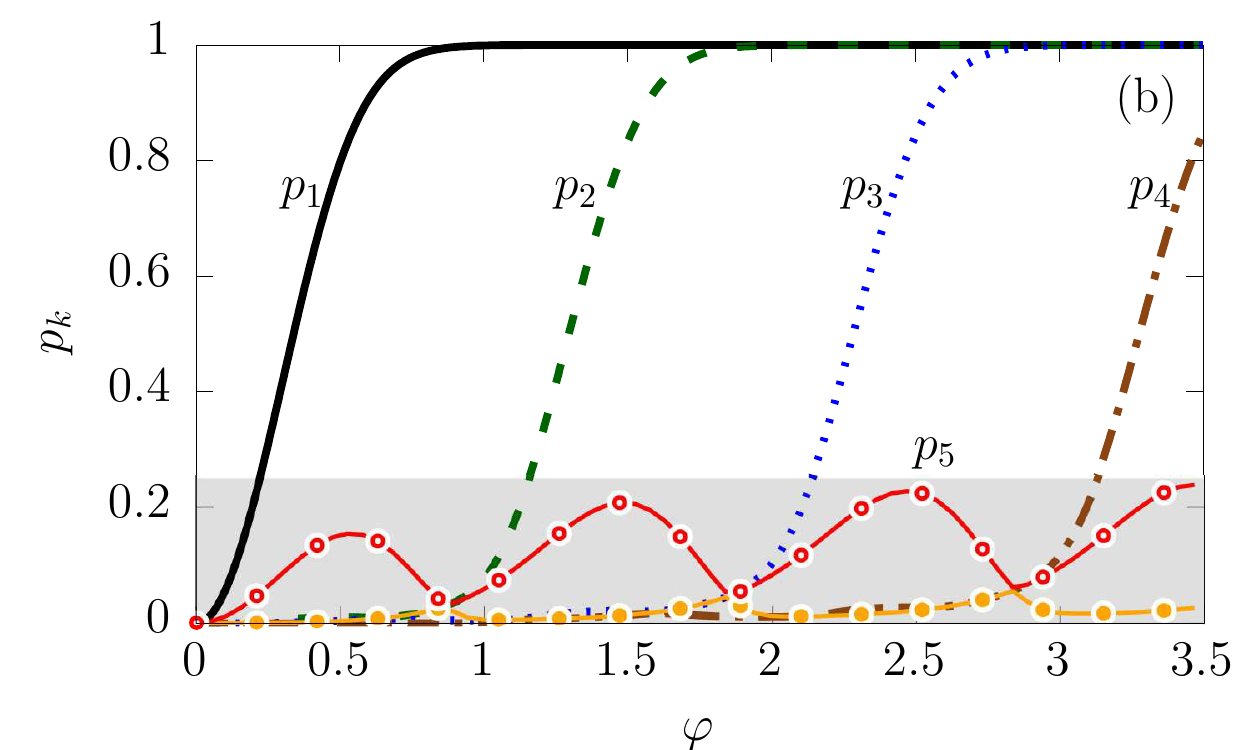}
  \includegraphics[width=6.5cm]{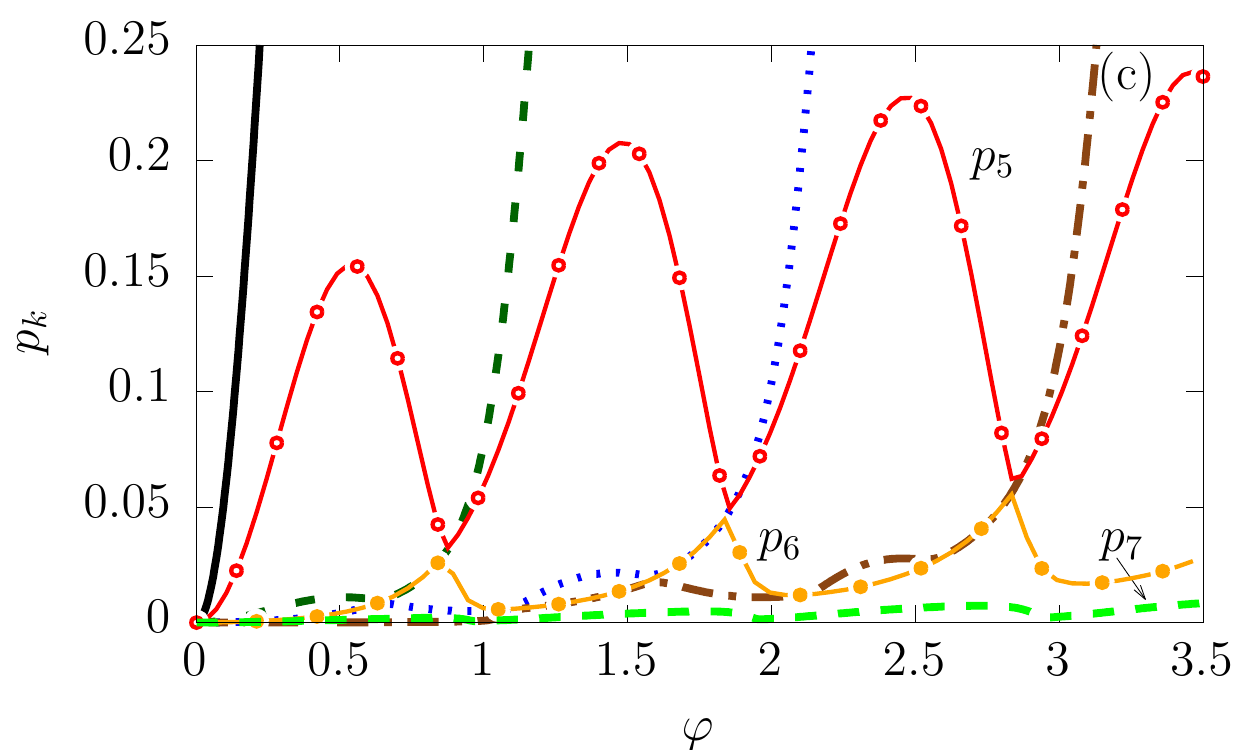}
  \caption{(Color online) (a) Schematic of the Square-profile pulse in time-domain. (b) Excitation probabilities as
    functions of the flux $\varphi$, with the width $W/t_0 = 1/8$. (c) Zoom-in plot of the grey regime.}
  \label{fig9}
\end{figure}

Triangular profile:
\begin{eqnarray}
 V_{p}(t) =
  \begin{cases}
    \frac{\pi\hbar}{2e} \frac{\varphi}{W^2}(2W-|t|)       &, \quad |t|<2W \\
    0  &,  \quad \text{otherwise.}
  \end{cases}
  \label{A1:eq2}
\end{eqnarray}

\begin{figure}[H]
  \centering
  \includegraphics[width=6.5cm]{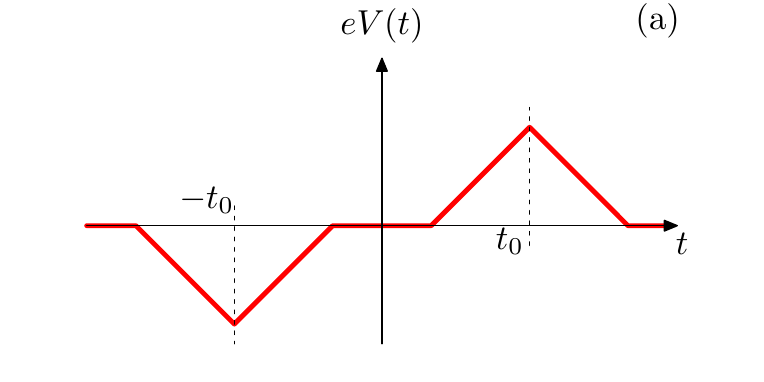}
  \includegraphics[width=6.5cm]{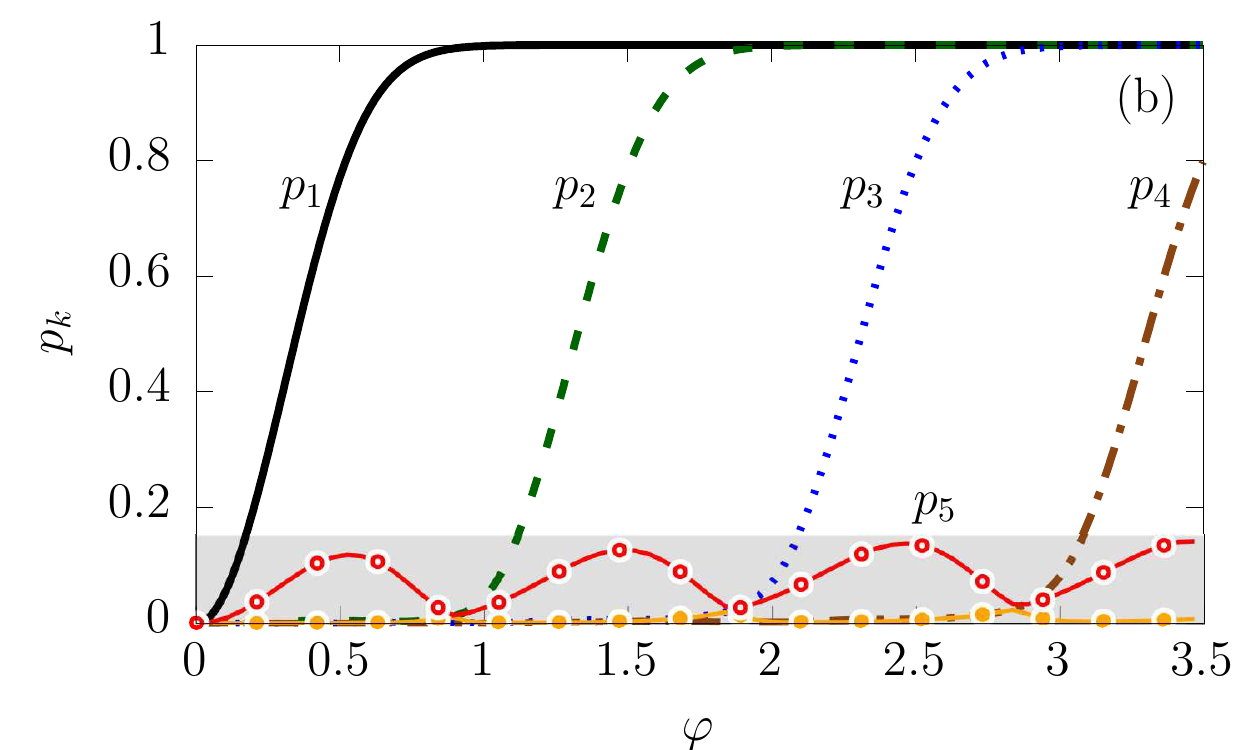}
  \includegraphics[width=6.5cm]{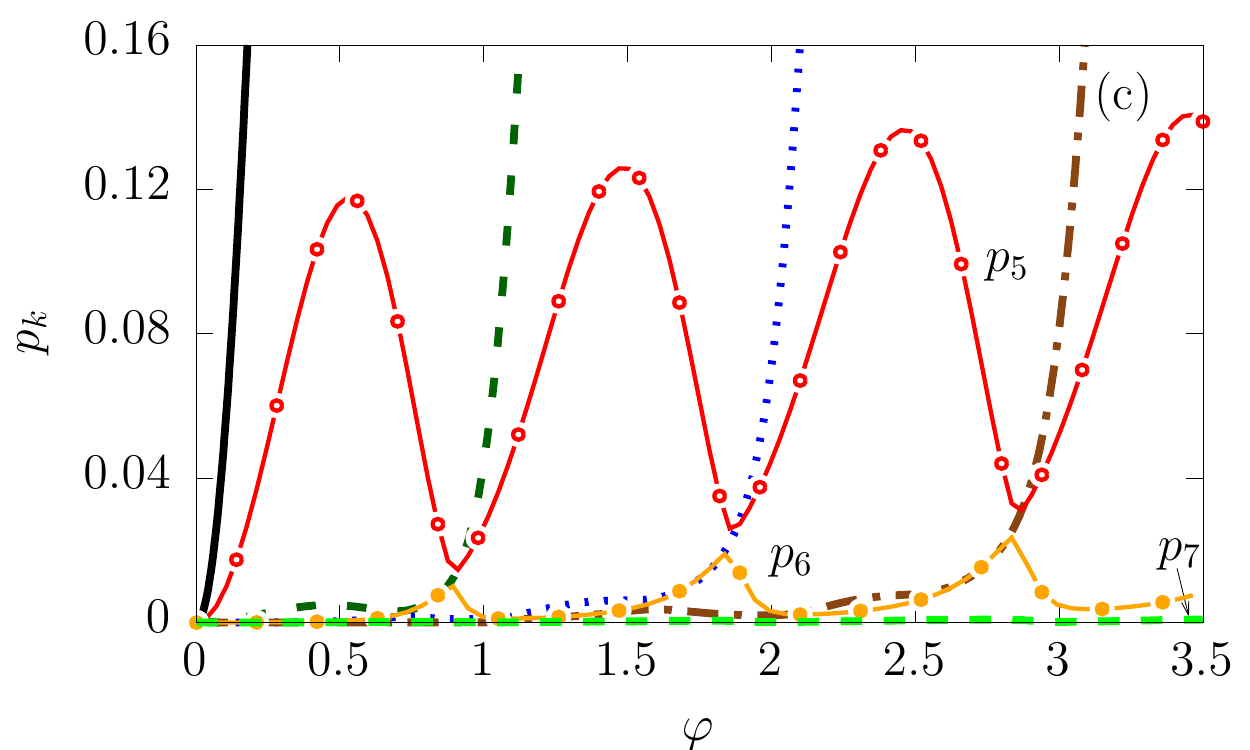}
  \caption{(Color online) (a) Schematic of the Triangular-profile pulse in time-domain. (b) Excitation probabilities as
    functions of the flux $\varphi$, with the width $W/t_0 = 1/8$. (c) Zoom-in plot of the grey regime.}
  \label{fig10}
\end{figure}

Parabolic profile:
\begin{eqnarray}
 V_{p}(t) =
  \begin{cases}
    \frac{3\pi\hbar}{2\sqrt{2}e} \frac{\varphi}{W}(1-\frac{t^2}{2W^2})  
    &, \quad |t|<\sqrt{2}W \\
    0  &, \quad \text{otherwise.}
  \end{cases}
  \label{A1:eq3}
\end{eqnarray}

\begin{figure}[H]
  \centering
  \includegraphics[width=6.5cm]{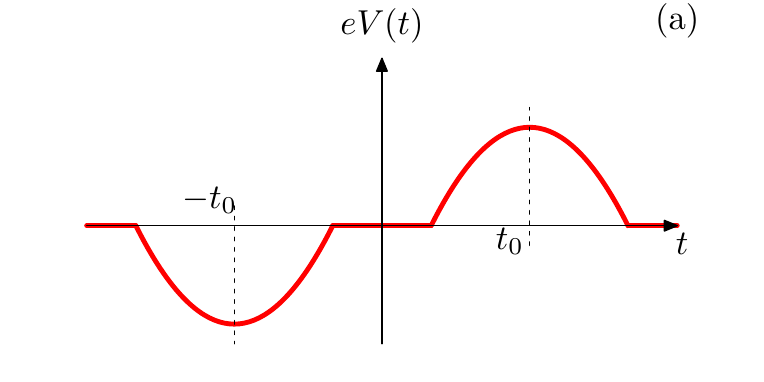}
  \includegraphics[width=6.5cm]{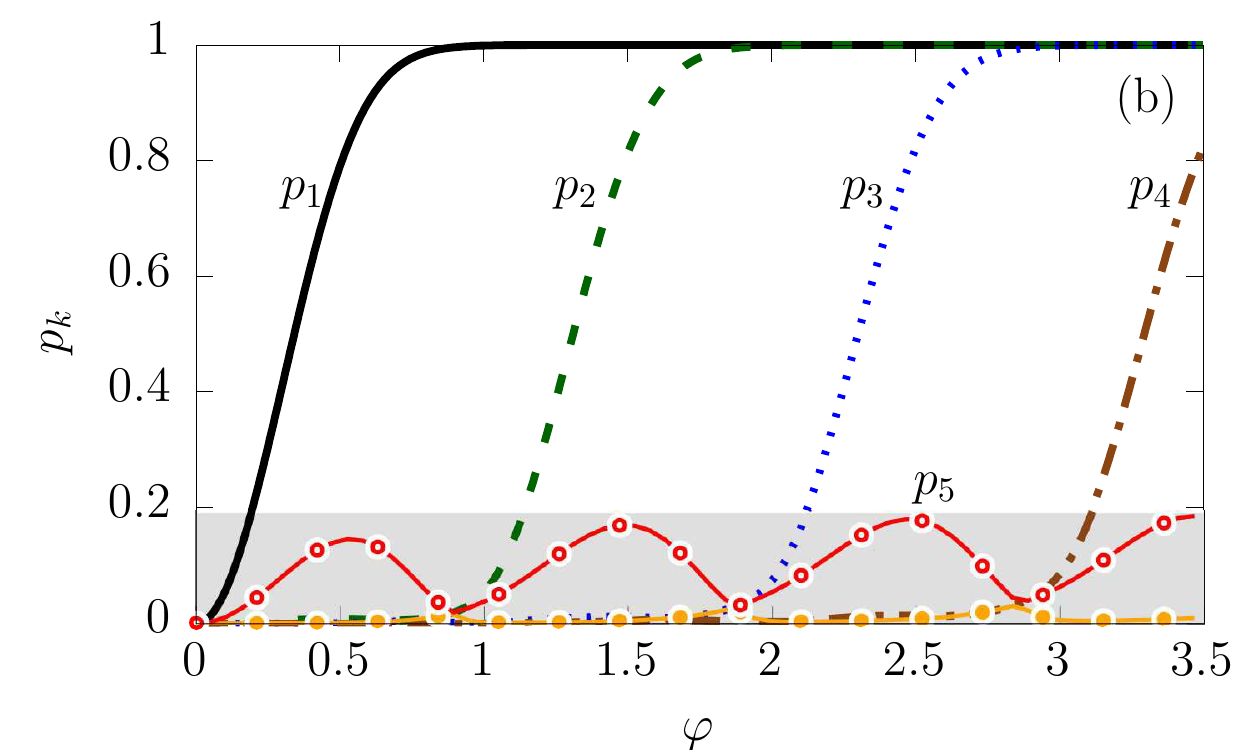}
  \includegraphics[width=6.5cm]{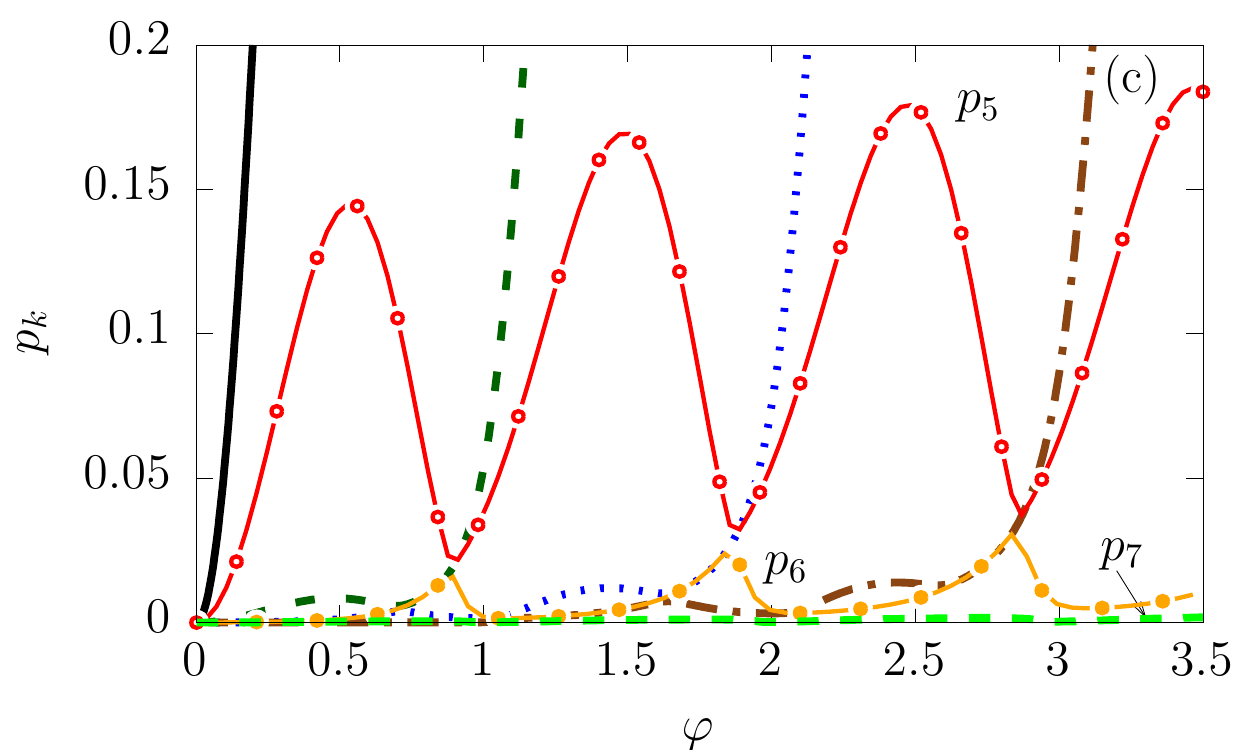}
  \caption{(Color online) (a) Schematic of the Parabolic-profile pulse in time-domain. (b) Excitation probabilities as
    functions of the flux $\varphi$, with the width $W/t_0 = 1/8$. (c) Zoom-in plot of the grey regime.}
  \label{fig11}
\end{figure}

\section{Effective binomial distribution of other profiles}
\label{app2}
In this appendix, we show the parameters $\alpha$ and $\bar{p}$ of the effective binomial distribution {\em without} and
{\em with} the contribution of the abnormal eh pairs for different profiles.

\begin{figure}[H]
  \centering
  \includegraphics[width=6.5cm]{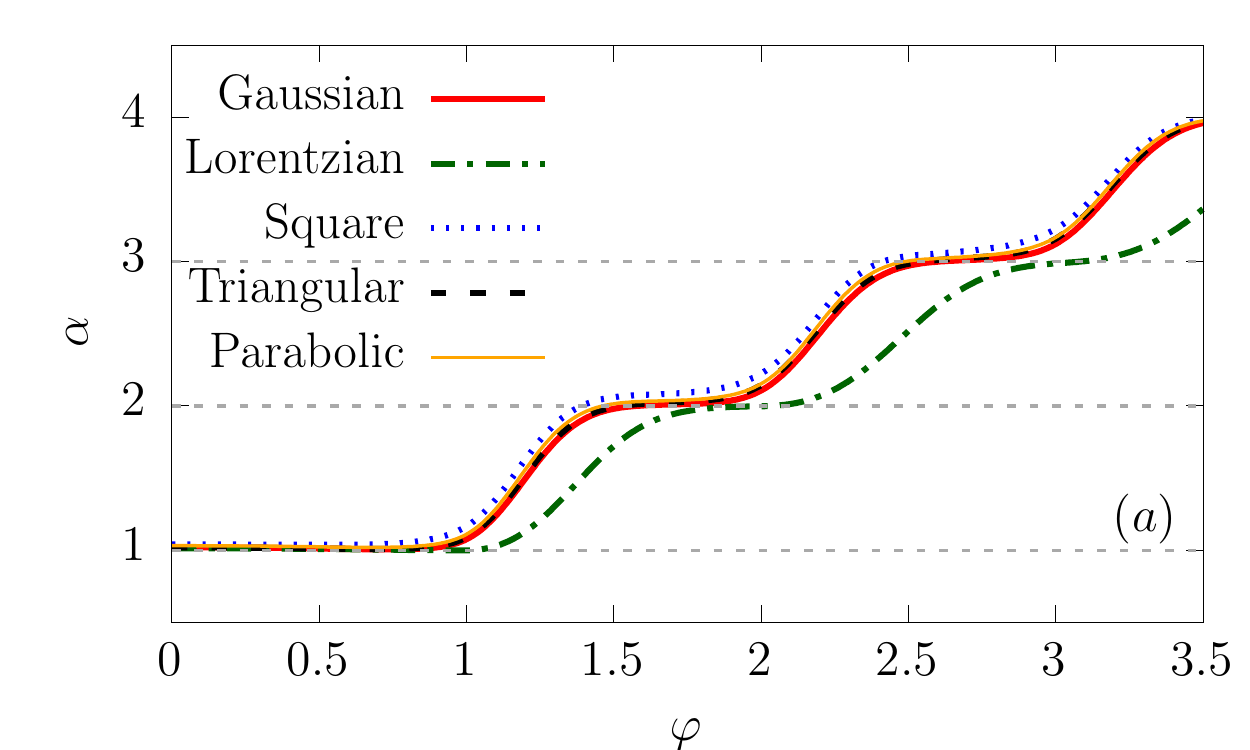}
  \includegraphics[width=6.5cm]{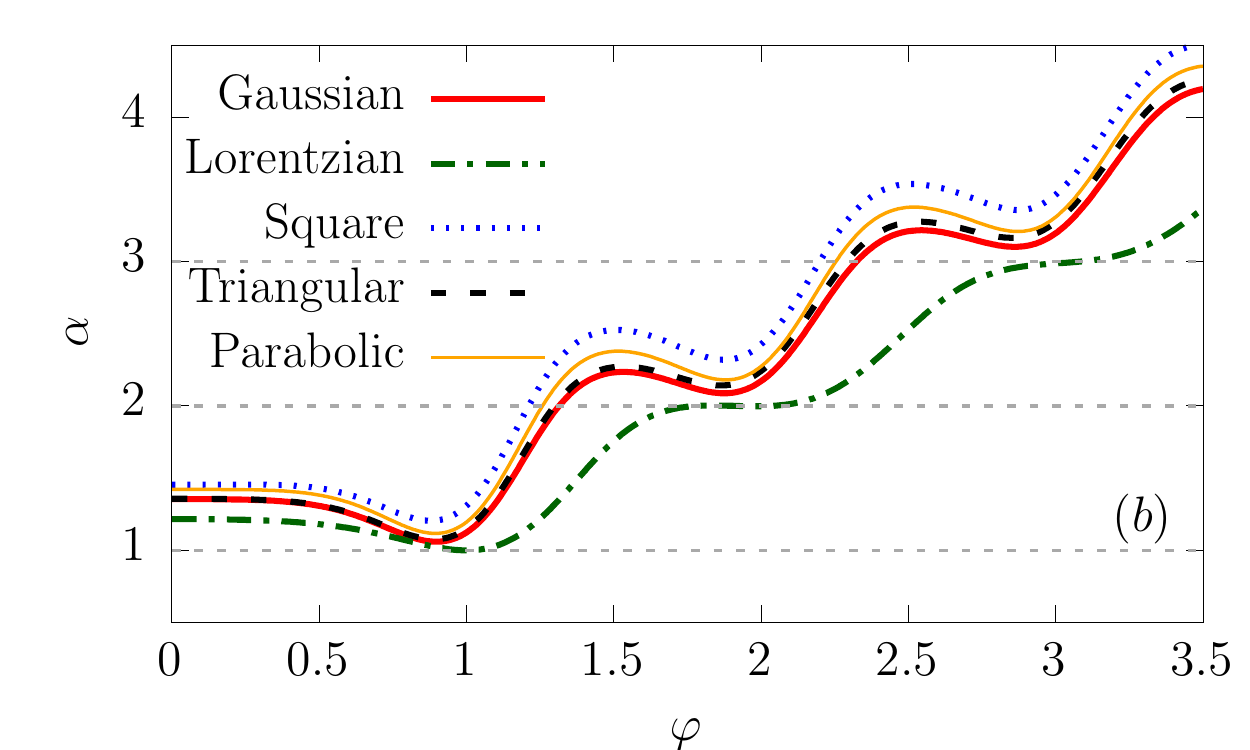}
  \caption{(Color online) The parameter $\alpha$ of the effective binomial distribution (a) {\em without} and (b) {\em
      with} the contribution of the abnormal eh pairs. The red solid, green dash-dotted, blue dotted, black dashed, and
    orange thin solid curves correspond to the Gaussian, Lorentzian, Square, Triangular and Parabolic profiles,
    respectively.}
  \label{fig12}
\end{figure}

\begin{figure}[H]
  \centering
  \includegraphics[width=6.5cm]{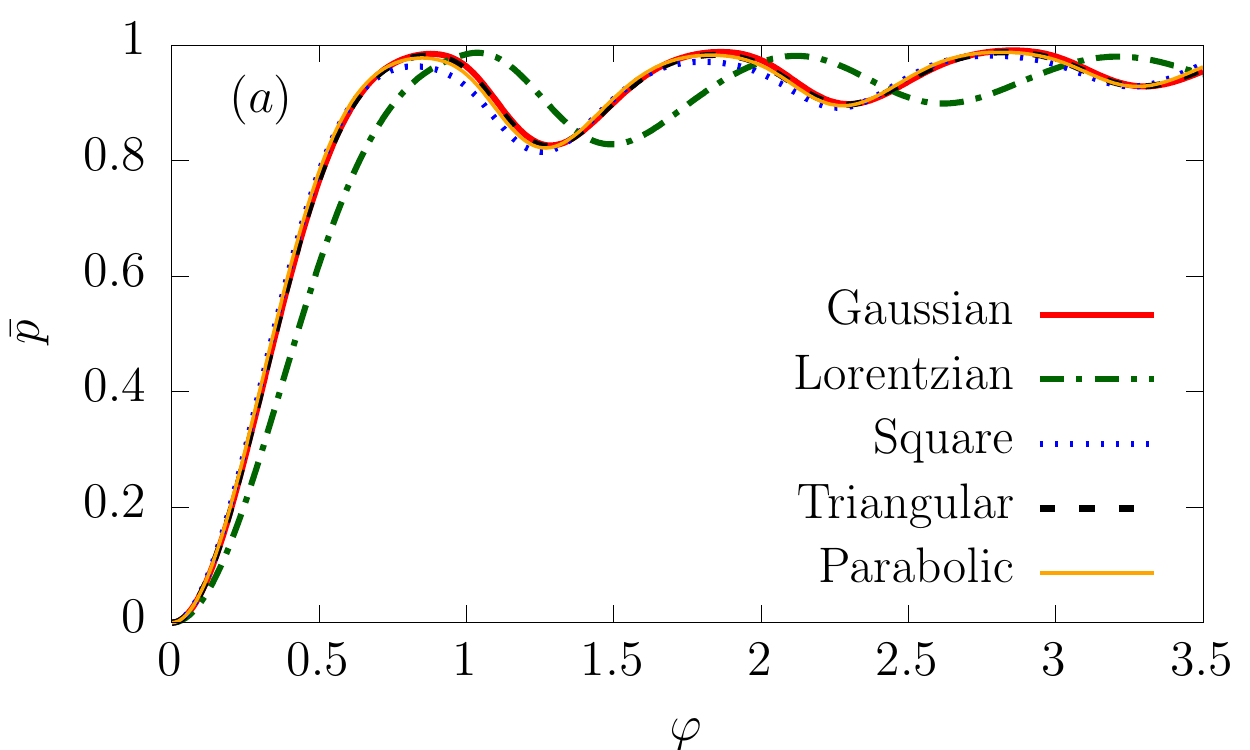}
  \includegraphics[width=6.5cm]{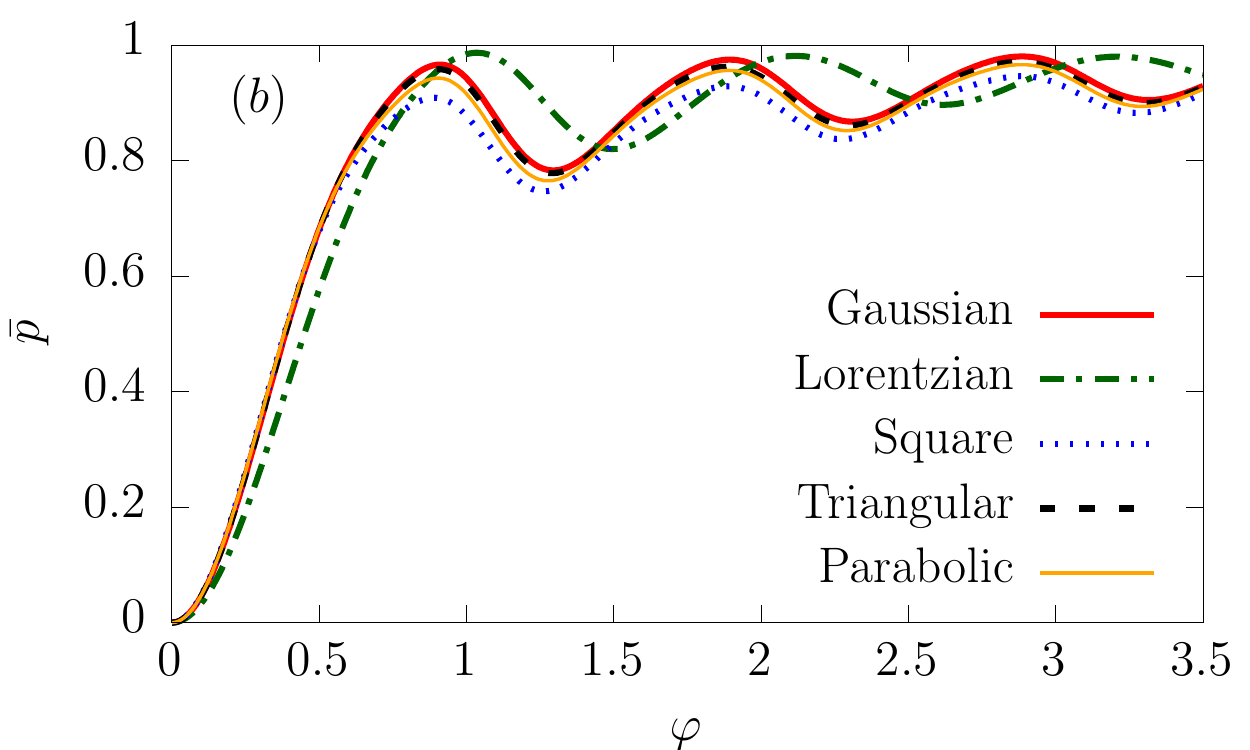}
  \caption{(Color online) The parameter $\bar{p}$ of the effective binomial distribution (a) {\em without} and (b) {\em
      with} the contribution of the abnormal eh pairs. The red solid, green dash-dotted, blue dotted, black dashed, and
    orange solid curves correspond to the Gaussian, Lorentzian, Square, Triangular, Parabolic profiles respectively.}
  \label{fig13}
\end{figure}

\bibliographystyle{apsrev4-1}

%

\end{document}